\documentclass[twocolumn]{aastex631}

\usepackage{nicefrac}
\usepackage{xcolor}

\newcommand{\comment}[1]{}
\newcommand{\lessthan}{\textless \space}
\newcommand{\greaterthan}{\textgreater \space}
\newcommand{\angstrom}{\AA \space}
\newcommand{\kms}{km s$^{-1}$}

\defcitealias{Opitsch+2018}{O18}
\defcitealias{Saglia+2018}{S18}

\shorttitle{M31 Stellar Chemodynamics}
\shortauthors{Gibson et al.}

\begin{document}

\title{The Chemodynamics of the Stellar Populations in M31 from APOGEE Integrated Light Spectroscopy}
\author[0000-0001-8203-6004]{Benjamin J. Gibson}
\author[0000-0001-6761-9359]{Gail Zasowski}
\author[0000-0003-0248-5470]{Anil Seth}
\author[0000-0003-1143-8502]{Aishwarya Ashok}
\author[0000-0002-7743-9906]{Kameron Goold}
\affiliation{Department of Physics and Astronomy, University of Utah, Salt Lake City, UT. 84112, USA}

\author[0000-0001-6320-2230]{Tobin Wainer}
\affiliation{Department of Physics and Astronomy, University of Utah, Salt Lake City, UT. 84112, USA}
\affiliation{Department of Astronomy, University of Washington, Box 351580, Seattle, WA 98195, USA}

\author[0000-0001-5388-0994]{Sten Hasselquist}
\affiliation{Space Telescope Science Institute, Baltimore, MD. 21218, USA}

\author[0000-0002-9771-9622]{Jon Holtzman}
\author[0000-0003-2025-3585]{Julie Imig}
\affiliation{Department of Physics and Astronomy, New Mexico State University, Las Cruces, NM. 88003, USA}

\author[0000-0002-3601-133X]{Dmitry Bizyaev}
\affiliation{Apache Point Observatory and New Mexico State University, P.O. Box 59, Sunspot, NM. 88349-0059, USA}
\affiliation{Sternberg Astronomical Institute, Moscow State University, Moscow, Russia}

\author[0000-0003-2025-3147]{Steven R. Majewski}
\affiliation{Department of Astronomy, University of Virginia, Charlottesville, VA. 22904-4325, USA}

\correspondingauthor{Benjamin J. Gibson}
\email{ben.gibson@utah.edu}

\begin{abstract}
We present analysis of nearly 1,000 near-infrared, integrated light spectra from APOGEE in the inner $\sim$7 kpc of M31.  We utilize full spectrum fitting with A-LIST simple stellar population spectral templates that represent a population of stars with the same age, [M/H], and [$\alpha$/M].  With this, we determine the mean kinematics, metallicities, $\alpha$ abundances, and ages of the stellar populations of M31's bar, bulge, and inner disk ($\sim$4-7 kpc).  We find a non-axisymmetric velocity field in M31 resulting from the presence of a bar.  The bulge of M31 is metal-poor relative to the disk ([M/H] = $-0.149^{+0.067}_{-0.081}$ dex), features minima in metallicity on either side of the bar ([M/H] $\sim$ -0.2), and is enhanced in $\alpha$ abundance ([$\alpha$/M] = $0.281^{+0.035}_{-0.038}$).  The disk of M31 within $\sim$7 kpc is enhanced in both metallicity ([M/H] = $-0.023^{+0.050}_{-0.052}$) and $\alpha$ abundance ([$\alpha$/M] = $0.274^{+0.020}_{-0.025}$).  Both of these structural components are uniformly old at $\simeq$ 12 Gyr.  We find the metallicity increases with distance from the center of M31, with the steepest gradient along the disk major axis ($0.043\pm0.021$ dex/kpc).  This gradient is the result of changing light contributions from the metal-poor bulge and metal-rich disk.  The chemodynamics of stellar populations encodes information about a galaxy's chemical enrichment, star formation history, and merger history, allowing us to discuss new constraints on M31's formation.  Our results provide a stepping stone between our understanding of the Milky Way and other external galaxies.
\end{abstract}

\section{Introduction \label{sec:intro}}
Roughly half of the stellar mass in the universe is contained in galaxies whose mass is within a factor of two of the Milky Way’s \citep[e.g.,][]{Baldry+2012, Driver+2022}.  Therefore, by studying the chemical abundances, dynamics, and their joint distributions (`chemodynamics’) of stars in these galaxies, we can further understand the physical processes that govern the behavior of baryonic matter.  We have a detailed understanding of the chemical makeup and kinematics of the stars in the Milky Way (MW), as our perspective from within the galaxy allows us to observe individual stars both photometrically and spectroscopically throughout it.  Surveys such as \textit{Gaia} \citep{Gaia_mission}, APOGEE \citep{Majewski+2017}, LAMOST \citep{LAMOST}, GALAH \citep{GALAH}, and several others, have observed up to billions of stars throughout the MW and provided us with detailed chemodynamical information for each of them.  When we break these stars down into groups with similar chemodynamics, we can identify sub-populations that likely have were born in different locations at different times \citep[e.g.,][]{Helmi+1999, Belokurov+2006, Helmi+2018, Myeong+2019, Koppelman+2019}.  This practice has made the MW the primary basis for understanding how MW-like spiral galaxies form and evolve.\\
\indent Unfortunately this perspective from inside the MW has some drawbacks.  For example, stars on the far side of the disk and bulge can often be obscured by gas and dust or other foreground stars.  Being in the plane of the disk complicates the classification of spiral arms and the bar, leading to different interpretations of galactic structures \citep[e.g.,][]{Lee+2015, Ness+Lang2016}.  It is also difficult to determine global properties of the MW such as total luminosity, color, mass-to-light ratio, bulge-to-disk mass ratio, etc. as we cannot see all of the MW in the same way we can see external galaxies \citep{Bland-Hawthorn+2016, Fielder+2021}.\\
\indent Luckily, we can work around some of these issues by studying our neighbor, M31, which is believed to be relatively similar to the Milky Way.  Despite its distance of $\sim785$ kpc \citep[e.g.,][]{M31Distance} and inclination of 77$^{\circ}$, we are able to observe the whole of M31's disk and bulge.  Like the MW, M31 is a large disk galaxy, though it features rings rather than spiral arms.  The stellar mass of M31's disk is $5.9\times10^{10}$ M$_\odot$ \citep{Yin+2009}, about 1.5$\times$ that of the MW's at $4\times10^{10}$ M$_\odot$ \citep{Sick+2015}.  M31's disk scale length \citep[5.76 kpc,][]{Dorman+2013} is about 2.2$\times$ larger than the MW's \citep[2.6$\pm$0.5 kpc,][]{Bland-Hawthorn+2016}.  These two galaxies are similar in metallicity \citep[e.g.,][]{PHAT_metal} and environment, being in the same galaxy group.  M31 has a bar, but it is much weaker and less distinct than the MW's \citep{Athanassoula+Beaton2006}, and M31's bulge is composite: $\nicefrac{1}{3}$ of its stellar mass is in a classical bulge and the other $\nicefrac{2}{3}$ is in a boxy/peanut-shaped (b/p) bulge \citep{Blana-Diaz+2017}.  The MW bulge is mostly b/p \citep{Wegg+Gerhard2013}.  The bulge-to-disk mass ratio for M31 is 0.43 \citep{Courteau+2011}, compared to 0.3 for the MW \citep{Bland-Hawthorn+2016}.\\
\indent The stellar populations of M31's bulge and disk have been the focus of numerous studies over the last 80 years.  \citet{Baade1944} barely resolved stars at the tip of the RGB in M31's bulge and concluded they were more similar to stars seen in the solar neighborhood of the MW, rather than the types of stars seen in globular clusters.  In the following decades, several resolved star studies concluded M31's bulge was made of old, metal-rich, population I stars \citep[e.g.,][]{Baum+Schwarzschild1955,Spinrad+Taylor1971,Wu+1980,Bohlin+1985,Walterbos+Kennicutt1987}.  \citet{Bica+Alloin+Schmidt1990} measured the metallicity of the bulge and found it to be super solar ([Z/Z$_{\odot}$] $\sim 0.3$).  The first element abundance measured in the bulge was [Mg/Fe] at 0.3 by \citet{Sil'chenko+1998}.\\
\indent Recently, resolved stellar photometry has become available across a vast extent of M31 \citep[e.g.,][]{PHAT, PAndAS}.  \citet{Olsen+2006} studied resolved NIR photometry in the bulge and disk, finding that the bulge is old ($\ge 6$ Gyr) and has solar metallicity.  They also note that the stellar ages are uniform throughout the bulge and inner disk.  The Panchromatic Hubble Andromeda Treasury \citep[PHAT,][]{PHAT} measured UV, optical, and IR photometry of 117 million individual stars in the bulge and northern half of M31.  The star formation history (SFH) of M31 recovered from this survey showed a majority of the stars in the disk are older than 8 Gyr, with a smaller component that has formed in the last 2-4 Gyr \citep{PHAT_SFH_old}.  The 5 kpc inner ring was also found to have ongoing star formation in the last 200 Myr \citep{PHAT_SFH_young}.  Spectroscopic kinematic measurements of 5800 PHAT stars shows a very clear age-velocity dispersion correlation in the disk, where older stars have a higher dispersion \citep{Dorman+2015}.  A negative metallicity gradient (-0.2 dex/kpc) was found from 4-20 kpc in the disk, as well as a slight metallicity enhancement along the bar from 3-6 kpc \citep{PHAT_metal}.\\
\indent Over the last decade or so, several studies have utilized simple stellar population models to analyze integrated light spectroscopy of M31's bulge and disk. 
 \citet{Saglia+2010} examined long slit spectra of the bulge region and found it to be uniformly old (12 Gyr), of solar metallicity, and to have an alpha abundance of [$\alpha$/Fe] $\sim 0.2$.  The very center is younger and slightly more metal-rich.  A study of LAMOST \citep{LAMOST} spectra in the inner 7 kpc agrees that the bulge is uniformly old, and the disk has solar metallicity and a negligible gradient \citep{Zou+2011}.\\
\indent To date, the most detailed integrated light studies of M31 come from \citet{Opitsch+2018}  (hereafter referred to as \citetalias{Opitsch+2018}), who examined optical integral-field unit (IFU) data of the inner bulge and parts of the disk.  They constructed detailed kinematic maps of this region and found strong evidence for a bar.  \citet{Saglia+2018} (hereafter referred to as \citetalias{Saglia+2018}) examined the same data to construct maps of the chemistry and age of the stellar populations.  They found the bulge to be old ($\sim$13 Gyr), metal-rich (roughly solar abundance) along the bar, and $\alpha$ enhanced (0.2-0.3 dex).  \citet{Gajda+2021} developed a made-to-measure model of M31 that matched the same data and found that the part of the bulge on either side of the bar was metal-poor relative to the bar and disk.  They called this pattern a metallicity ``desert".\\
\indent In this paper, we analyze high resolution near-infrared integrated light spectra from APOGEE \citep{Majewski+2017} of M31 out to a radius of $\sim$7 kpc\footnote{We measure radii as deprojected onto the disk of M31 using a position angle of 37.7$^\circ$, an inclination angle of 77$^\circ$, and a distance of 785 kpc.  All quantities measured in kpc have been deprojected in this manner.} (see Figure \ref{fig:M31data}).  Our data cover the entire bar of M31 as well as the inner disk and inner ring at 5 kpc.  This data set is unique compared to previous integrated light spectroscopic studies of M31.  No other data set has complete coverage of the inner regions, as most span a few fields in the central regions, or have continuous coverage that doesn't extend quite so far out.  These data are also higher spectral resolution than previous studies by nearly an order of magnitude, which allows us to measure the average kinematics of the stellar populations with unprecedented accuracy.\\
\indent APOGEE's near-infrared (NIR) coverage contains spectral features from several $\alpha$ elements \citep{APOGEE_line_list}, which can be used to trace the SFH and chemical evolution of stellar populations.  The NIR is significantly less affected by dust attenuation than the optical and is sensitive to RGB and AGB stars.  This allows us to clearly observe and more completely sample the stellar population.  Moreover, APOGEE's measurements of $\sim$650,000 stars in the MW have been leveraged to create a library of single stellar population (SSP) model spectra \citep{Ashok+2021} with the same spectral resolution and wavelength range.  This library is unique to our study, affording us an advantage over other studies, as they may have to navigate discrepancies between their data and models, such as differences in spectral resolution, that we do not.  Therefore, we can analyze our M31 data with our unique models to enable direct, consistent comparison between the stellar populations of the MW and M31.\\
\indent In this paper we present maps of the mean kinematics, abundances, and ages of the stellar population of M31.  These maps cover, for the first time, the entirety of the bulge, bar, inner disk, and 5 kpc ring.  Section \ref{sec:obs_data} describes the APOGEE observations and data processing, Section \ref{sec:models} details our SSP models, and Section \ref{sec:methods} describes the data binning and fitting analysis.  The results are presented and discussed in Section 4, and the summary and conclusion are in Section 5.

\section{Observations and Data Processing \label{sec:obs_data}}
\indent Data for this project was taken as part of the Apache Point Observatory Galactic Evolution Experiment \citep[APOGEE,][2023, \textit{in prep}]{Majewski+2017}. APOGEE, part of SDSS-III \citep{Eisenstein+2011} and SDSS-IV \citep{Blanton+2017}, is a NIR (1.5-1.7 $\mu$m), high resolution ($R\sim22,500$), multi-object spectroscopic survey.  APOGEE observed $\sim$650,000 stars in the Milky Way (MW), as well as individual stars in nearby dwarf galaxies and integrated light spectra from extragalactic GCs and nearby galaxies \citep{Zasowski+2013, Zasowski+2017, Beaton+2021,Santana+2021}.  Data analyzed in this paper were taken using the APOGEE spectrograph \citep{Wilson+2019} on the 2.5 meter SDSS telescope at Apache Point Observatory \citep{Gunn+2006}.  APOGEE also includes data taken a second APOGEE spectrograph on the du Pont Telescope at Las Campanas Observatory \citep{Bowen+Vaughan1973}.\\
\indent The standard data processing pipeline for APOGEE is outlined in \citet{Nidever+2015}, and the latest data release (DR17) was described in \citet{Abdurro'uf+2022}.  The APOGEE Stellar Parameter and Chemical Abundances Pipeline \citep[ASPCAP;][Holtzmann et al. 2023, \textit{in prep}]{Perez+2016, Nidever2021} determines stellar parameters and $\sim$20 elemental abundances for stars in the APOGEE data set.  Note that ASPCAP was used to develop the models described in Section \ref{sec:models}, not for abundance determination of our M31 data.\\
\indent Throughout this paper we use `fiber position' to indicate a location in M31 that has been observed.  Each fiber position is observed multiple times, and we will refer to each individual observation as a `visit' (see \citet{Zasowski+2013} Section 2.1 for more details).

\subsection{Observations \label{subsec:observations}}
\begin{figure}[h!]
\includegraphics[width=3.35in]{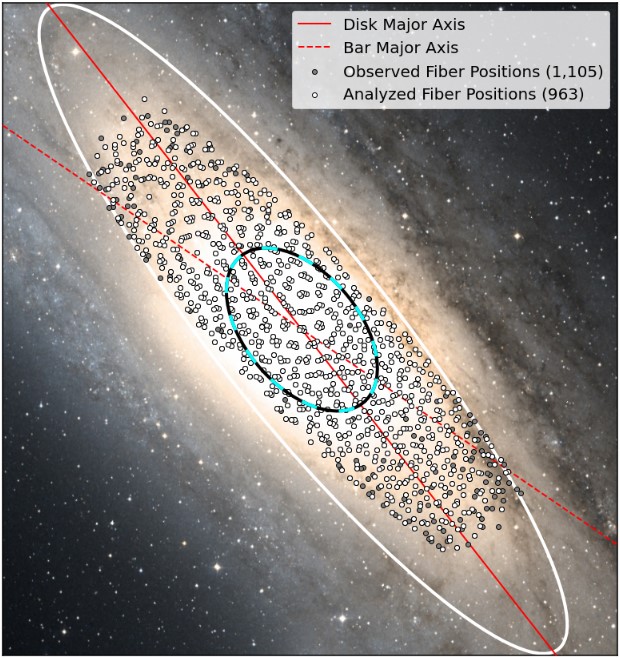}
\caption{APOGEE fiber positions overlaid onto the DSS optical image of M31 (Credit: Infrared Science Archive at IPAC and California Institute of Technology).  The solid red line is the disk major axis \citep[37.7$^{\circ}$,][]{M31_disk_MA}; the dotted red line is the bar major axis \citep[55$^{\circ}$,][]{Beaton+2007}.  The black and cyan ellipse denotes the boundary of the bulge region, and the white ellipse shows 7 kpc from the center deprojected onto the disk. \label{fig:M31data}}
\end{figure}
The M31 observations were performed as part of an ancillary science program of APOGEE \citep{Zasowski+2017}.  Five sets of targets were designed to observe a total of 1,105 locations.  These designs were drilled into plug plates for fiber optic cables that gather the light from an object and transmit it to the APOGEE spectrograph.  One plate with an M31 design can be seen in Figure \ref{fig:plugplate} in Appendix \ref{sec:plate}.  The fiber placement was designed to be as evenly spaced as possible, but was subject to several constraints.  APOGEE fibers are 2" in diameter, and the collision distance between fibers was 72".  The design of the first plate included the very center of M31 and allocated fibers outward from there.  The other plates were designed to start 10" from the center and their fiber positions were no closer than 10" from fiber positions in any other plate.  Fiber positions avoided bright sources identified with 2MASS \citep{2MASS} imagery, as these were assumed to belong to the Milky Way.  Sky and telluric calibration fiber positions were determined by standard APOGEE routines (see Sections 5.1 and 5.2 of \citet{Zasowski+2013} for more details).  They are located outside the disk of M31 and are roughly uniformly dispersed across a circle $\sim1.75^{\circ}$ in radius from the center of M31. All of the calibration was done using standard APOGEE procedures (see Sections 6.2 and 6.3 of \citet{Nidever+2015}).\\ 
\indent Figure \ref{fig:M31data} shows the distribution of fiber positions on an optical image of M31.  Positions filled with gray are not included in the final analysis (see Section \ref{subsec:cutfibs}).  For the remainder of this paper, the term ``bulge" will refer to the region inside the black and cyan ellipse in Figure \ref{fig:M31data}.  This was chosen to match the region where the bulge makes up \textgreater50\% of the total luminosity of the galaxy according to \citet{Courteau+2011} (see Section \ref{subsubsec:decomposition}).  With this definition, the ``bulge" dominates the light to $\sim$1.80 kpc along the disk major axis and $\sim$5.56 kpc along the disk minor axis, due to the inclination of M31's disk.  We note that radii measured in kpc are deprojected onto the disk of M31.\\
\indent These observations extend to a radius of $\sim$7 kpc along the bar major axis.  Each fiber position was observed ten times for roughly 66 minutes each time.  There are 275 fiber positions in the bulge region; the other 830 are in the inner disk region, roughly from 3-7 kpc from M31's center.  Our fibers cover roughly 150 square kpc of M31's bulge and disk, for an average density of $\sim$7.4 fibers per square kpc.\\ 

\subsection{Data Processing}
The APOGEE pipeline is optimized for data of individual stars, not integrated light of external galaxies, which often have low SNR and high velocity dispersion.  To account for this, we modified the final step of the standard APOGEE pipeline: the combining of several visits of each fiber position into a single spectrum.  The entire process is described in \citep{Nidever+2015}.  We refined this final step as described below to adequately identify and alleviate errors (from skylines, detector persistence, etc) while preserving as much of our data as possible.\\
\indent The data gathered for an object observed by APOGEE is condensed into an apStar file \citep{Nidever+2015}.\footnote{APOGEE documentation: \url{https://www.sdss.org/dr17/irspec/}}  These files contain target flux spectra, sky flux spectra, target flux uncertainty, and pixel bitmasks for each visit to a fiber position, as well as each of these for a single combined spectrum.\footnote{apStar Data Model: \url{https://data.sdss.org/datamodel/files/APOGEE_REDUX/APRED_VERS/stars/TELESCOPE/FIELD/apStar.html}}  The apStar files will only include a visit if it is used to create the final combined spectrum, excluding data with very low SNR and poor radial velocity (RV) determination.\footnote{\url{https://www.sdss.org/dr17/irspec/radialvelocities/}}  These choices were optimized for stellar spectra, and exclude some of our low SNR and high dispersion data.  For this reason, the standard data release products contain spectra for only 877 fiber positions\footnote{These standard apStar files can be found at \url{https://data.sdss.org/sas/dr17/apogee/spectro/redux/dr17/stars/apo25m/} in either ANDR3 or ANDR3shift}.  We worked with the APOGEE data team to create custom apStar files that removed the SNR cutoff and the RV and barycentric corrections.  We received 1,082 custom apStar files that contain seven to ten visits each.  These custom apStar files, as well as the final data set outlined in Section \ref{subsec:cutfibs}, are available upon request and will be made publicly available as part of a future publication.\\

\subsubsection{Masking Individual Visit Spectra \label{subsubsec:Masking}}
\indent The APOGEE spectrograph is made of three CCD chips covering different wavelength ranges.  There is no data gathered for the wavelengths between the chips.  To avoid these chip gaps, we split each spectrum into three separate arrays, one for each chip.  Each chip is then masked, combined, and stored individually.\\
\indent To combine the visit spectra into one final spectrum, we need to identify pixels with unreliable fluxes and significant sky emission.  To do this, we first take a 500 pixel running median of the object flux spectrum for each visit to use as a continuum (this same strategy is used by the APOGEE data reduction pipeline).  We then calculate $\Delta$Flux = $|$flux - continuum$|$, as well as S/O = $\frac{\textrm{sky flux}}{\textrm{object flux}}$ for each pixel.  $\Delta$Flux was chosen as a means to simply remove the pixels that are farthest away from the continuum, which are generally, but not always, due to sky emission.  S/O was chosen as a means to also identify pixels with high sky flux.  We then calculate the 85th percentile for $\Delta$Flux and S/O and mask pixels that fall above this line for both statistics.  Pixels flagged by one statistic are often, but not always, flagged by the other.  A percentile cutoff was chosen as it will remove the most affected $\sim$15\% of pixels from a spectrum regardless of the visit's SNR.\\
\indent We also mask pixels in each visit spectrum according to the visit-level PIXMASK flags.\footnote{\url{https://www.sdss.org/dr17/algorithms/bitmasks/}, see APOGEE\_PIXMASK: APOGEE pixel level mask bits}  Specifically, we mask out pixels with bits 0, 1, 2, 3, 4, 5, 6, 7, 8, 9, 10, and 14 set.  These bits identify pixels affected by, among other things, medium to high levels of persistence, bad calibration frames, cosmic rays, or saturation.\\
\indent Upon visual inspection, a number of our spectra contained what appeared to be a prominent emission feature centered near 15930 \AA. This feature was identified as an artifact resulting from decreased sensitivity in a small region of the detector;  the dome flat shows almost no flux between 15890 to 15960 \angstrom for fibers with IDs between 20 and 30.  Since dome flats are removed by dividing, a low amount of flux from the flat will cause a spike in the object flux in the affected region.  We mask out this region in affected fibers.\\
\indent In summary, our final visit-level pixel mask flags pixels affected by severe sky emission, several issues identified by the APOGEE pixel masks, and one region of low detector sensitivity in the green chip.  This amounts to masking $\sim$25\% of our available pixels.  We note that these issues are specific to our data and not indicative of larger issues with the APOGEE data processing pipeline.
\subsubsection{Combining Masked Visit Spectra \label{subsubsec:stacking}}
The products from the previous section are masked visit spectra of each fiber position. 
 For a given fiber position, each visit's chip spectra are psuedo-continuum normalized using a new 500 pixel running median that ignores the pixels masked in Section \ref{subsubsec:Masking}.  The visits are then combined using an inverse variance weighted average:
\begin{equation}
    F = \frac{\Sigma \frac{f_i}{\sigma_i^2}}{\Sigma \frac{1}{\sigma_i^2}} \textrm{ ; } \sigma = \frac{1}{\Sigma \frac{1}{\sigma_i^2}}
\end{equation}
where $F$ and $\sigma$ are the combined pixel flux and uncertainty, respectively, and $f_i$ and $\sigma_i$ are the visit pixel continuum-normalized flux and uncertainty, respectively.\\
\indent If a pixel is masked in at least half of the available visits, then it is masked out in the final combined spectrum.  These pixels are assigned a flux value equivalent to the average continuum flux at that pixel and are given a very large uncertainty, we used 10$^{15}$ in flux units.  This combining process is the same process used in the apStar files\footnote{\url{https://www.sdss.org/dr17/irspec/spectral_combination/}} \citep{Nidever+2015}, with the primary distinctions being that we remove masked pixels from the combined spectrum and leave our final spectra continuum normalized.\\

\subsection{Determining the Final Data Set \label{subsec:cutfibs}}
For our final data set that we consider in the rest of the paper, we removed fiber positions from our sample for which we were unable to determine reliable kinematics from the final combined spectrum.  Specifically, we removed fiber positions that had inconsistent velocities from chip to chip or that had unrealistically high velocity dispersions.  The majority of these fiber positions are in the outskirts, and have low SNR spectra that complicate the measurement of their kinematics.  We performed the preliminary full-spectrum fitting routine with \textsc{ppxf} \citep{pPXF} defined in Section \ref{subsec:fitting}, which matches our data with the model spectral templates described in Section \ref{sec:models}; the kinematics are allowed to vary freely from chip to chip in this procedure.  From these fits we determined the three templates with the highest weight used in the \textsc{ppxf} fit (weights normalized across each chip).  Next, each chip was fit to each of the three templates individually and we determined which combination of templates gave the smallest variation in velocities from chip to chip.  We identified 95 fiber positions with RV variations \greaterthan50 \kms, then visually inspected each of these spectra and kept 7 fiber positions that had otherwise consistent kinematics and RV variations only slightly above 50 \kms.  We also identified 22 fiber positions from the 95 that appeared to have prominent spectral features but nevertheless gave inconsistent kinematics.  These 22 spectra were fit again using the MCMC routine described in Section \ref{subsec:fitting} and we kept nine whose kinematics converged well.\\
\indent In the end, 32 fiber positions had no visits with SNR above one, so these were removed from the sample.  We cut 79 fiber positions due to high RV variations between chips.  Three other fiber positions had dispersions in one or more chips \greaterthan220 km/s, and we do not expect values this high in the disk of M31, so they were cut.  Another 3 fiber positions were cut that had $\chi^2$ values, which are used to determine the goodness-of-fit, above 5 in the blue and red chips, or above 8 in the green chip (green chip fits are, on average, worse).  Lastly, we cut 2 more fiber positions that were unintentionally centered on GCs.  We identified these fiber positions by their low dispersion (\lessthan 60 km/s) and verified the GC using imaging from the PHAT survey \citep{PHAT}.  Our final sample consists of 963 fiber positions, or 89\% of our initial 1,082 fiber position sample (see Figure \ref{fig:M31data}).

\subsection{Velocities from Opitsch et al. 2018 \label{subsec:opitsch}}
Preliminary testing showed that we could only reliably recover the velocity curve seen in M31 from individual fiber positions with an empirical signal-to-noise ratio (eSNR, see Section \ref{subsec:fitting}) of $\gtrsim$ 20.  We also found that an eSNR of $\sim$ 60 was needed to determine abundances.  Therefore, we bin fiber positions together in the outer disk to increase their eSNR, which requires reliable velocity measurements for each fiber position.  We adopt mean stellar radial velocity data from \citetalias{Opitsch+2018} for shifting and combining binned data (see Section \ref{subsec:abundancebin}).\\
\indent \citetalias{Opitsch+2018} studied the stellar kinematics of M31 using the VIRUS-W optical integral-field unit (IFU) spectrograph mounted on the 2.7m telescope at the McDonald Observatory \citep[$R\sim9,000$,][]{VIRUS-W}.  The \citetalias{Opitsch+2018} data cover M31's bulge as well as azimuthal strips, or `spokes', of the disk out to 5.3 kpc along the disk major axis.  We take \citetalias{Opitsch+2018}'s measured line-of-sight velocities and LOESS smooth them using the LOESS Package \citep{loess}, which implements the Locally Weighted Regression algorithm of \citet{LOWESS} to adaptively smooth over observational errors and unveil the data's underlying structure.  We smoothed the data with a degree of 1, fraction of 0.5 (both are parameters in the LOESS package code that help control the smoothness), and included the error bars.  This algorithm can extrapolate onto locations that lie beyond the spatial extent of the data; some of our fiber positions lie outside of the bulge region measured by O18 (see Figure \ref{fig:viruscomp}).  The smoothed map was evaluated at our fiber positions to get a radial velocity measurement for each; we call this set of LOESS-smoothed velocity measurements the ``O18 velocities".  These were compared to velocities we measured for spectra we Voronoi binned (in the same manner as Section \ref{subsec:abundancebin}) to a target eSNR of 20.  Subtracting the former from the latter we found a bias of -0.65 \kms\space and a scatter of $\sim$19 \kms.  The velocities published in \citetalias{Opitsch+2018} have errors of 3-4 \kms, which are comparable to our single fiber position errors.  A plot comparing our velocities to those of \citetalias{Opitsch+2018} can be found in Appendix \ref{sec:velcomp}.\\

\section{SSP Models \label{sec:models}}
\subsection{A-LIST}
The APOGEE Library of Infrared SSP Templates \citep[A-LIST;][]{Ashok+2021} is a library of SSP template spectra generated from APOGEE stellar spectra.  A-LIST spans a wide range of population ages, metallicities, and $\alpha$ abundances.  This library of spectral templates was designed specifically to analyze IL spectra from APOGEE.\\
\indent We only utilize templates with luminosity fraction \greaterthan0.32, which is a measure of the completeness of the APOGEE sample for a given age, metallicity, and $\alpha$ abundance, as recommended by Section 4.1.1 in \citet{Ashok+2021}.  We also restrict our template sample to have [M/H] $\ge$ -1.0 and Age $\ge$ 7 Gyr.  The method we use to interpolate between A-LIST templates (described next) can be simplified by restricting our parameter space, since there is an age-metallicity degeneracy in the lowest metallicities.  Moreover, we do not expect to find the average population to be more metal-poor or younger in the bulge and inner disk of M31 \citep[e.g.,][]{Olsen+2006, Saglia+2018}.

\subsection{A-LIST Interpolation}
\begin{figure}
\includegraphics[width=3.35in]{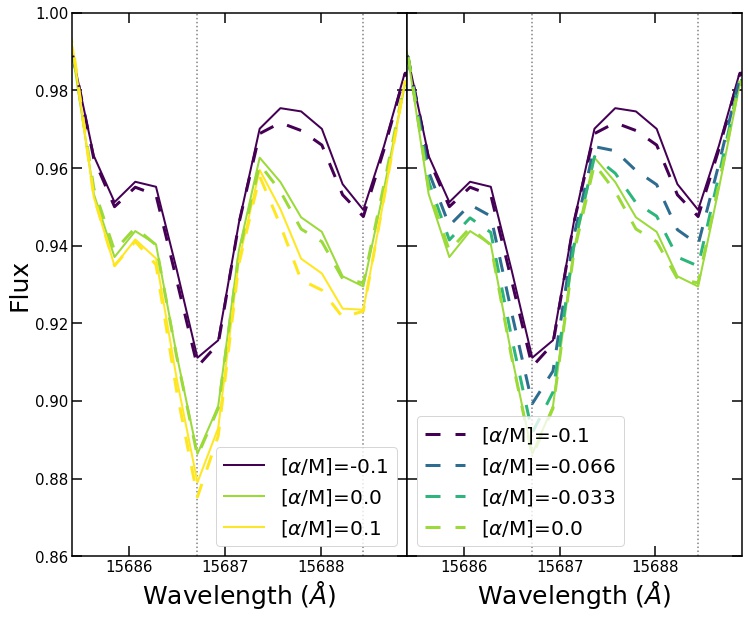}
\caption{Comparison of spectral templates in \textit{The Cannon} model interpolation (dashed) and the original A-LIST templates (solid) for age = 9 Gyr, [M/H] = 0.2 populations.  There are two spectral features shown here:  the left dotted vertical line marks the wavelength of an Fe I absorption feature, and the right dotted vertical line marks the location of a Ti I absorption feature.  The left panel shows how our interpolation routine accurately reproduces the shape of the A-LIST templates; the right panel shows the interpolation between discrete A-LIST templates.\label{fig:speccomp}}
\end{figure}

We interpolate between the discretely gridded A-LIST templates to use them with the MCMC fitting method described in Section \ref{subsec:fitting}. We adopt \textit{The Cannon} \citep{TheCannon1, TheCannon2} for this interpolation.  \textit{The Cannon} is a data-driven method to transfer stellar parameters and abundances (labels) from a training set of spectra to a broader data set.  The labels for the training set are determined through independent methods.  \textit{The Cannon} models the flux in each pixel as a linear combination of the labels.  In our case, we train the model on our sample of A-LIST spectra using [M/H], [$\alpha$/M], and $\log_{10}$(Age) as labels.  This allows us to generate model spectra continuously across our parameter space.\\
\indent This interpolation routine has the added benefit of fixing a handful of cases where $\alpha$-element lines do not deepen linearly with increasing $\alpha$ abundance due to the luminosity fraction and $\Delta T_{eff}$ limits of the library.  Additionally, \textit{The Cannon} training procedure allows the user to add weights to pixels.  \textit{The Cannon} procedure ensures that upweighted pixels will be recreated accurately relative to downweighted ones.  We upweight pixels that show the strongest variation of flux with $\alpha$ abundance, $\frac{dF}{d\alpha},$\footnote{In Figure \ref{fig:speccomp} the flux at the bottom of the deeper line for the [$\alpha$/M] = 0.0 template is 0.805.  For the [$\alpha$/M] = 0.3 template, the flux here is 0.771.  Therefore the slope $\frac{dF}{d\alpha} = \frac{0.085-0.771}{0.0-0.3} = -0.113$, and the weight W = 1.13.} holding age and metallicity constant.  In particular, we weight a fraction of the pixels by:
\begin{equation}
    W = -c \frac{dF}{d\alpha}
\end{equation}
where W is the weight and c is 10.  We apply this to the 10\% of pixels with the steepest slopes.  The other 90\% of pixels are given a weight of 1.

\begin{figure}
\includegraphics[width=3.35in]{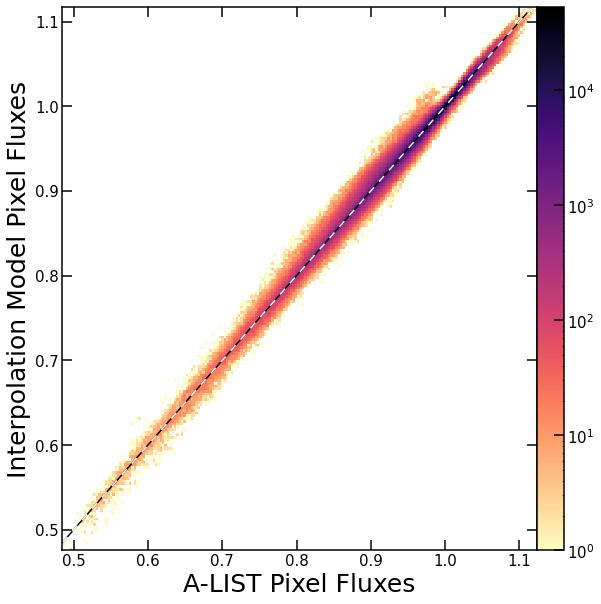}
\caption{Comparison of pixel fluxes in the interpolation and the original A-LIST templates.  The colorbar indicates the number of pixel fluxes at a given point.  The small number of outliers to the upper left of (1,1) are due to the low metallicity, low $\alpha$ abundance templates in our sample.  There are not many stars with this chemistry in the MW, and as such these templates are more sensitive to fluctuations from star to star.  Therefore, these templates are not necessarily as representative of their populations as templates with higher metallicities or $\alpha$ abundances, and they are affected the most by our interpolation routine.\label{fig:fluxcomp}}
\end{figure}

We tested many models, varying the multiplier and percentile, as well as the polynomial order and regularization of \textit{The Cannon}.  The regularization parameter controls the amount to which the flux determined for a given pixel is influenced by the flux of surrounding pixels.  Following \citet{Ashok+2021}, we analyzed APOGEE IL spectra of M31 GCs with [M/H] $\ge$ -1.0.  We found the model that best matched the original A-LIST spectra and recreated the results from \citet{Ashok+2021} was an unregularized third order polynomial with a multiplier of 10 on the slopes of the top 10\% steepest pixels.\\
\indent As can be seen in the left panel of Figure \ref{fig:speccomp}, the original A-LIST templates are well reproduced by our interpolation routine.  From the right panel one can see that we are able to reasonably model the flux for abundances between the discrete A-LIST grid points.  Figure \ref{fig:fluxcomp} shows the distribution of the flux in each pixel of all the interpolated models vs. the A-LIST models.  We show that the interpolated model is a good recreation of A-LIST, with a median absolute deviation of 0.0013 normalized flux units between the two.

\section{Binning and Fitting \label{sec:methods}}

\subsection{Binning \label{subsec:abundancebin}}
\indent To bin adjacent fiber positions together, we first shift the spectra to a common wavelength array determined by the average of the O18 velocities of the fiber positions in each bin.  We then stack the spectra together using the routine outlined in Section \ref{subsubsec:stacking}, and calculate the eSNR with the first two steps of our General Fitting Procedure (Section \ref{subsec:fitting}).\\
\indent We determine bin members by Voronoi binning in radius and position angle using the code VorBin \citep{VorBin}.  Voronoi binning allows us to increase our eSNR to a target of 70 while maintaining as much spatial resolution as possible.  An added desire for our bins is a small spread in the intrinsic velocity dispersions of the spectra in a bin.  Velocity dispersion trends closely with radius, so we multiply the radius by six so VorBin thinks the fiber positions are further away radially than they actually are.  This produces bins as radially narrow as possible.  Lastly, we set the zero of position angle to the southern disk semi-minor axis so that we can have bins that cross over the major axis.\\
\indent We made some manual edits to the final binning results to even out the eSNRs, and in the end we have 164 bins with an eSNR $\gtrsim$ 60.  The final bins are shown as the colored polygons in Figure \ref{fig:mosaic}; the fiber positions included in each bin are the black and white dots inside each polygon.  In Section \ref{subsec:choices} we discuss the robustness of this method to the choices involved.

\subsection{General Fitting Procedure \label{subsec:fitting}}
We utilize full spectrum fitting via \textsc{ppxf} \cite[Penalized Pixel-Fitting,][]{pPXF} to determine the kinematics, abundances, and ages of the stellar populations in M31.  \textsc{ppxf} is a reduced $\chi^2$ minimization software that shifts and broadens template spectra (in our case, the interpolated A-LIST model) to fit data.  We assume that the stellar population making up our IL spectrum has the line of sight velocity and velocity dispersion found by the \textsc{ppxf} fitting procedure. The light-weighted mean age, metallicity, and $\alpha$ abundance of the population at that fiber position is the age, metallicity, and $\alpha$ abundance of the best fitting interpolated SSP template.  That template is determined by an MCMC routine that searches the entire parameter space for the set with the maximum likelihood, $L = \nicefrac{-\chi^2}{2}$, using the \textbf{emcee} software package \citep{emcee}.\\
\indent The fitting procedure is as follows:
\begin{enumerate}
    \item \textsc{Mask the following pixels in each spectrum: (in addition to \ref{subsubsec:Masking})}
    \begin{itemize}
    \item[-] The 250 pixels closest to the red end of each chip, and the 100 pixels closest to the blue end of each chip.  This is to avoid the chip gaps present in our templates that get blue-shifted into the fit by \textsc{ppxf}.
    \item[-] Pixels in each spectrum within 2.5 \AA\space of skylines with an intensity larger than 0.5 from \citet{Rousselot+2000}.\footnote{The line list can be found here: \url{https://www2.keck.hawaii.edu/inst/nirspec/ir_ohlines.dat}}
    \item[-] Pixels in the following ranges that were repeatedly found to be poorly fit by the A-LIST templates: 15360-15395, 16030-16070, and 16210-16240 \AA\space (See Section \ref{subsubsec:misfit}).
    \end{itemize}
    \item \textsc{Calculate eSNR:} Using \textsc{ppxf}, fit each chip individually with an mdegree of 40, which uses a multiplicative polynomial of order 40 to represent the continuum of the spectrum. This value was chosen as it was found to produce fits with no structure in the residuals.  Fit the data with templates interpolated at the discrete A-LIST grid points.  Use the $\chi^2$ from this fit to calculate the rank-based estimate of the standard deviation of the residuals over the uncertainties, $q$:
    \begin{equation}
        \sigma_G = .7413 ( q_{75} - q_{25} )
    \end{equation}
    where $q_{75}$ and $q_{25}$ are the upper and lower quartiles of $q$.  Then, multiply the uncertainty arrays by $\sigma_G$ to make the $\chi^2$ value for this fit equal one.  Use these scaled uncertainties for the final fit in step 3 and to calculate the eSNR of the spectrum.
    \item \textsc{Perform the fit:} Initialize \textbf{emcee} routine bounds and other parameters.  Set the walker bounds as follows: Velocity Dispersion between 60 and 200 \kms, [M/H] between -1.0 and 1.0, [$\alpha$/M] between 0 and 0.5, and Age between 7 and 12 Gyr.  Initialize walkers randomly between these bounds.  Set the mdegree for the \textsc{ppxf} fit to 2.  Run the \textbf{emcee} routine using StretchMove\footnote{StretchMove is the default \textbf{emcee} "move": an algorithm that dictates how the walkers move through the parameter space from iteration to iteration} with the desired number of kinematic and stellar population parameters; this can be four or five (see next paragraph).  Use 30 walkers for 200 iterations and a burn-in period of 125 iterations.\footnote{200 iterations surpasses the autocorrelation time.  Our burn-in period is larger than half the number of iterations due to our uninformative priors.}  Take the final results for each parameter as the median values in the ensemble chain after the burn-in period.
\end{enumerate}

\indent For bins comprised of more than one fiber position, we fit the spectra with a 4D MCMC routine to determine the mean dispersion, metallicity, $\alpha$ abundance, and age of the stellar population.  The RV is fixed to be the eSNR weighted mean of the O18 velocities of the fiber positions in the bin.\\
\indent For bins that contain just one fiber position, we fit the spectra with a 5D MCMC routine to determine the velocity in addition to the four parameters above.  For these fits, the velocity walkers are initialized in a uniform distribution $\pm$15 \kms\space around the O18 velocity for that fiber position.\\
\begin{figure}
\includegraphics[width=3.35in]{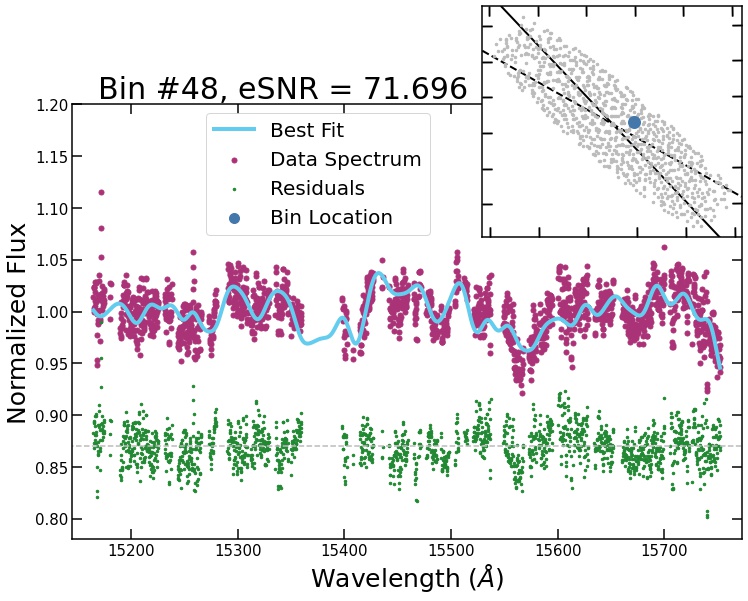}
\caption{An example of a fit to our data following the procedure described in Section \ref{subsec:fitting}.  The magenta dots show the data spectrum, with the best fit shown in light blue.  The residuals to the fit are the green dots.  The plot in the upper right shows the location of the bin. 
 \label{fig:datafit}}
\end{figure}
A fit to a bin near the Voronoi target eSNR of 70 is shown in Figure \ref{fig:datafit}; just the blue chip is shown.  Bin 48 consists of a single fiber due west of the center of M31.  The $\chi^2$ of this fit was 1.99.  The measured radial velocity is -367 \kms\space and the dispersion is 126 \kms.

\subsection{Measurement Error Determination \label{subsec:errorbars}}
The errors quoted in Section \ref{sec:results} are determined through a jackknife resampling procedure where data points are systematically removed from the sample and the rest are reanalyzed to determine the variations between data points \citep{jackknife}.  For our data the final spectrum for single-fiber-position bins is made by stacking visit spectra, whereas the final spectrum for multi-fiber-position bins is made by shifting and stacking the spectra for fiber positions in that bin.  We vary these constituent spectra in our jackknife process to determine the errors.  We calculate the jackknife-determined errors and an error scaling term, which is the median of the ratio of jackknife-determined errors to MCMC-determined errors, for bins with five or more constituent spectra.  For all other bins we multiplied the MCMC-determined errors by the scaling term to get the quoted error on each measurement.  In general our jackknife-determined errors are larger by a factor of 1-5 than the MCMC-determined errors, with no strong correlation between eSNR and error size.  Jackknife-determined errors measure the intrinsic variation in the sample, rather than the MCMC-determined errors which are statistical, hence the former being larger.  The median scaling used for each parameter was 2.13 for velocity, 2.43 for dispersion, 2.47 for metallicity, 1.93 for $\alpha$ abundance.\\

\section{Results and Discussion \label{sec:results}}

\begin{figure*}
\centering
\includegraphics[width=.8\textwidth]{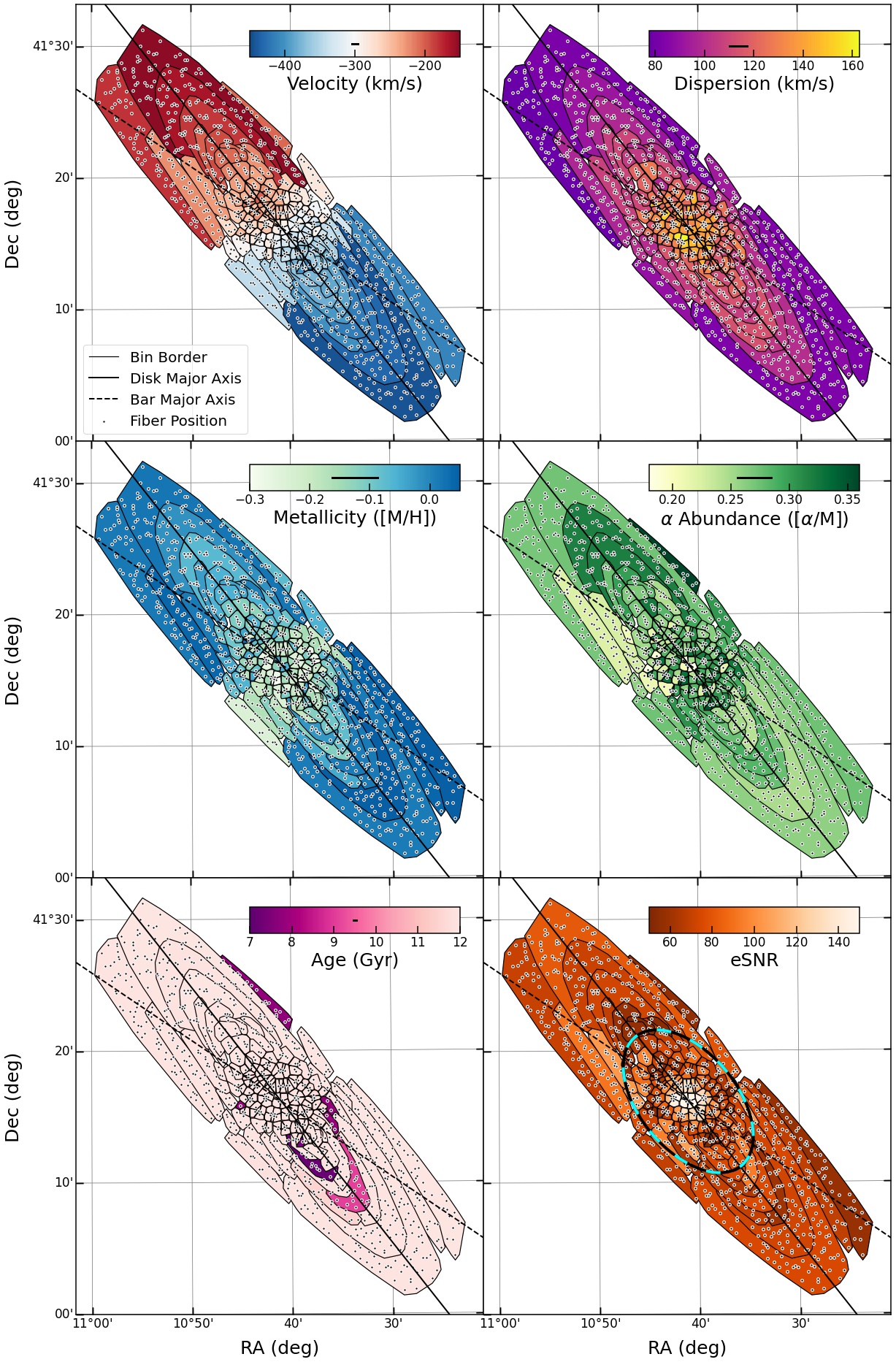}
\caption{\label{fig:mosaic}Maps of the mean kinematics, abundances, age, and eSNR for the stellar populations in the bulge and disk of M31.  Bins, the polygons outlined by thin black lines, are colored by the value of the parameter found by the MCMC routine outlined in Section \ref{subsec:abundancebin}.  The black and blue dashed ellipse in the eSNR plot is the region defined as the ``bulge".  Black lines in the color bars indicate the median error size on that parameter for all bins.  Fiber positions are the black and white dots, and the disk and bar major axes are indicated with the solid and dashed diagonal lines, respectively.}
\end{figure*}

\subsection{Mean Trends in the Bulge and Disk \label{subsec:trends}}
The primary results of this work are presented in Figure \ref{fig:mosaic}.  This figure shows the light-weighted average kinematics, chemistry, and ages of the stellar populations in the bulge and inner disk.  The bottom right panel of Figure \ref{fig:mosaic} shows the eSNR of each bin.  Median errors for each parameter are shown as black lines in the colorbars in Figure \ref{fig:mosaic}.\\
\indent In Sections \ref{subsubsec:velocity} to \ref{subsubsec:ages} we will discuss the mean trends for each parameter in Figure \ref{fig:mosaic}.  Section \ref{subsec:nucleus} will discuss the central five fiber positions that likely host significant nuclear populations.  Section \ref{subsec:gradients} describes our results for abundance gradients in the bulge.  We compare our results to the literature in Section \ref{subsec:litcomp} and examine other analysis choices we investigated in Section \ref{subsec:choices}.

\subsubsection{Radial Velocity \label{subsubsec:velocity}}
\begin{figure}
\includegraphics[width=3.35in]{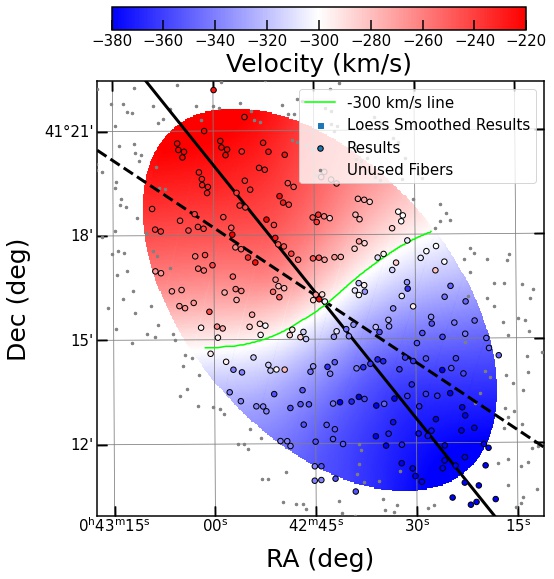}
\caption{\label{fig:loessvel}LOESS-smoothed radial velocity in the bulge.  Fiber positions used in the smoothing are indicated by black-outlined circles, colored by the value of the parameter used in the smoothing.  For velocity we smooth using a fraction of .5 and a degree of 1.  The lime line on the left-side panel indicates where the smoothed velocity equals -300 \kms, showing the `s' shaped velocity profile that is an effect of M31's bar.  As in Figure \ref{fig:mosaic}, the solid and dashed black lines indicate the disk and bar major axes, respectively.}
\end{figure}

The top left panel of Figure \ref{fig:mosaic} shows our radial velocity results for M31, which have typical errors of 3.8 \kms.  Our results are consistent with a systemic velocity of $-300\pm4$ \kms\space for M31 \citep{M31velocity, M31Distance}.  In Figure \ref{fig:loessvel} we LOESS smooth these results in the bulge.  The raw data is shown as colored circles with black outlines, and the smoothing is shown as the red, white, and blue background.  The lime line is a twist in the radial velocity profile along the direction of the bar.  This is a well known bar effect \citep[e.g.,][]{Stark+2018} and has previously been seen by \citetalias{Opitsch+2018}.  See Section \ref{subsubsec:barbulge} for further discussion. 

\subsubsection{Velocity Dispersion \label{subsubsec:dispersion}}
Results for velocity dispersion are shown in the top-right panel of Figure \ref{fig:mosaic}.  We show that the bulge has a dispersion \textgreater130 \kms\space that decreases roughly linearly ($\sim$-10 \kms\space kpc$^{-1}$) along all position angles.  The inner disk of M31 has a velocity dispersion $\leq$100 \kms.  Our radial velocity dispersion map peaks at $\sim$160 \kms\space off-center to the south-east.  The dynamics of the very center are complex \cite[e.g.,][]{Lockhart+2018}, and our results here are discussed more in Section \ref{subsec:nucleus}.

\subsubsection{Metallicity}
\begin{figure}
\includegraphics[width=3.35in]{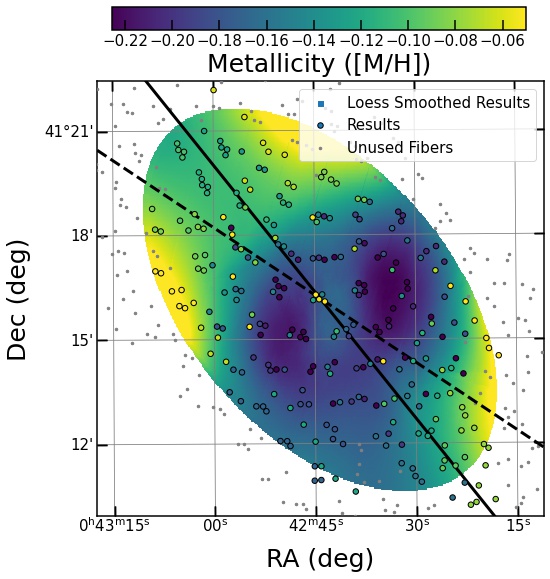}
\caption{\label{fig:loessmet}LOESS-smoothed metallicities in the bulge, as in Figure \ref{fig:loessvel}.  The metallicity shows a clear depression on either side of the disk major axis and bar.  The five nuclear fiber positions show enhancement of roughly .1 dex relative to the surrounding fiber positions.  The parameters used for smoothing the metallicity are fraction=0.2 and degree=2.  As in Figure \ref{fig:mosaic}, the solid and dashed black lines indicate the disk and bar major axes, respectively}
\end{figure}

Our metallicity results are presented in the left middle panel of Figure \ref{fig:mosaic}.  The bulge is generally metal-poor (median [M/H] = -0.149 dex); the inner disk shows near-solar metallicity (median [M/H] = -0.006).  The nucleus is enhanced relative to the rest of the bulge (median [M/H] = -0.055).\\
\indent We find a clear pattern of metal deficiency in the bulge perpendicular to the bar major axis, which is shown in Figure \ref{fig:loessmet}.  This structure was uncovered by \citet{Gajda+2021} as so-called ``metallicity deserts".  \citetalias{Saglia+2018} determined that the bar of M31 is metal enhanced relative to the off bar regions, but we do not find as clear of a signature.  We attribute this to a lack of spatial resolution sufficient to entirely negate the effects of stochasticity.  We examine the metallicity gradients in Section \ref{subsec:gradients}.

\subsubsection{$\alpha$ Abundance}
\begin{figure}
\includegraphics[width=3.35in]{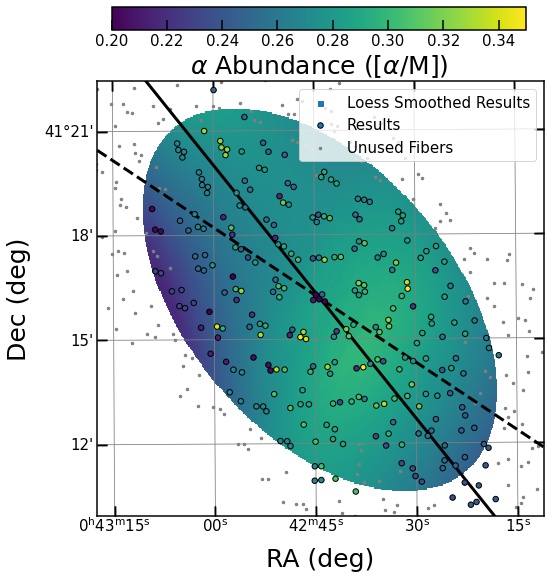}
\caption{\label{fig:loessalph}LOESS-smoothed $\alpha$ abundances in the bulge, as in Figure \ref{fig:loessvel}.  The bulge is uniformly enhanced relative to the disk by roughly .1 dex.  The central fiber position has an abundance of .063$\pm$.008 dex, and the other four nuclear fiber positions are also deficient.  The region to northeast is also $\alpha$-poor and is possibly the end of the bar.  The parameters used for smoothing the $\alpha$ abundance are fraction=0.5 and degree=2.  As in Figure \ref{fig:mosaic}, the solid and dashed black lines indicate the disk and bar major axes, respectively}
\end{figure}

The right middle panel of Figure \ref{fig:mosaic} shows our map of the $\alpha$ abundance in M31.  The bulge of M31 is enhanced in $\alpha$ abundance with a median [$\alpha$/M] = 0.281.  The results in this region exhibit statistically significant variations in the $\alpha$ abundance of 2.5 to 3 times our jackknife-determined errors.  These errors accurately quantify observational uncertainties, but don’t account of stochastic variation in the stellar population from point to point.  These statistical fluctuations do not appear to trace a coherent spatial pattern.  This is consistent with the bulge having no $\alpha$ abundance substructure, but rather having a spread in $\alpha$ abundance that is evident from stochastic variations in the fraction of light from stars of different $\alpha$ abundance that are in each fiber.\\
\indent The  nucleus is more $\alpha$-poor than the rest of the bulge, but still super-solar (median [$\alpha$/M] = 0.179), and the central fiber position has [$\alpha$/M] = 0.006.  The disk is also enhanced (median [$\alpha$/M] = 0.274), but shows more clear spatial substructure.  The LOESS smoothed map of the bulge is shown in Figure \ref{fig:loessalph}.  Amongst the features visible is lower $\alpha$ abundances along the minor axis and asymmetry along the bar major axis; a clear signature of this $\alpha$-poor region to the north-west is also seen by \citetalias{Saglia+2018}.  It could be that this is the end of the bar protruding out from the bulge into the disk, however an equivalent feature is not seen as strongly on the opposite side of the bulge.  Analysis of the data with symmetric binning across the disk major axis was unable to produce a similar deficiency on both sides of the bar.  Dust lanes in the region could obscure the stellar populations to the south-west and increase the apparent enhancement, but further analysis of this asymmetry is beyond the scope of this work.

\subsubsection{Stellar Age \label{subsubsec:ages}}
The bottom left panel of Figure \ref{fig:mosaic} shows our results for the age of the stellar populations in M31.  We find the bulge to be uniformly 12 Gyr old.  Our model grid extends only to 12 Gyr, so our results are a lower bound.  In the disk we find most bins to be old as well, with a few exceptions.  We note that the A-LIST models are not particularly sensitive to age, as ages for spectra used to create A-LIST are from astroNN \citep{astroNN}, which trains a neural network on asteroseismic data to determine ages from spectra that are accurate to $\sim30-35\%$.  Additionally, two models with mono-abundance and ages only off by 1 Gyr are very similar (median flux differences below .001).  Therefore, we trust the bins with ages \textgreater11 Gyr; the median error on these ages is 0.31 Gyr. Bins with ages \textless11 Gyr have elevated errors (median 4.35 Gyr) and higher $\chi^2$s, as the MCMC routine cannot locate a maximum likelihood age for these few bins.  Analysis of these bins with age fixed to 12 Gyr generally decreased metallicity and $\alpha$ abundance, but these results remain in line with expectations (see Section \ref{subsubsec:fixedages} for more details).  We identified three bins with ages \textless8 Gyr and $\chi^2$ \textgreater3.5 whose fixed-age $\chi^2$ was greatly improved and will utilize those results for our analysis.

\subsection{Nuclear Regions \label{subsec:nucleus}}
The nuclear star cluster (NSC) of M31 dominates the light out to 5" ($\simeq.02$ kpc) \citep{Kormendy+Bender1999,Peng2002}, so our central fiber position is sensitive to only NSC stars.  We have four other fiber positions from 11-12" ($\simeq.045$ kpc) that are likely also sensitive to nuclear stars.  Thus we call these five fiber positions the ``nuclear fiber positions".  The stellar populations in this region are distinct from the bulge and inner disk populations.  We find these fiber positions to be metal enhanced ([M/H]$\simeq$-0.05) and $\alpha$ deficient ([$\alpha$/M]$\simeq$0.18) relative to the rest of the bulge.\\
\indent The central 4 pc of M31 were investigated by \citet{Lockhart+2018} who show that the dynamics of this region are complex.  They find a quickly rotating eccentric disk here with asymmetric velocity magnitudes on either side.  They find the dispersion to be peaked at 381$\pm$55 \kms\space 0.13" from the center, though this high-dispersion area is only $\sim$0.2" in diameter.  APOGEE fibers have an on-sky diameter of 2", implying our result for this fiber position should be the average value of their data.  They find that a larger area is red-shifted from -300 \kms\space than is blue-shifted.  Additionally, the majority of their dispersion plot is near 150 \kms.  This lines up with our result for the central fiber position, which has a radial velocity of $-225.56^{+0.486}_{-0.459}$ \kms\space and a velocity dispersion of $151.91^{+0.364}_{-0.307}$ \kms.  This indicates the NSC of M31 is towards the sun slower than the rest of M31.
\indent \citetalias{Opitsch+2018} found a ring of increased dispersion (up to $\sim$180 \kms) a few hundred arcseconds ($\sim$1 kpc) from the nuclear regions, which have a lower dispersion (around 150 \kms). \citet{Saglia+2010} show that the velocity dispersion reaches a minimum around 5" from the center before rising again further out.  This is consistent with our analysis, which indicates the peak dispersion being off-center by a few hundred arcseconds.  We would likely see a ring of increased dispersion too with better spatial sampling.  Both our results and those of \citetalias{Opitsch+2018} are subject to stochasticity.  \citetalias{Saglia+2018} show an increase in metallicity to [Z/H]=0.35 in the very center of M31 but no significant decrease in $\alpha$ abundance.  The central 5" hosts a mostly old stellar population \citep[7-13 Gyr][]{Saglia+2010}.\\
\indent We find the NSC of M31 to be old ($\simeq$12 Gyr) and to have a radial velocity of -225.56$\pm$3.23 \kms, velocity dispersion of 156.91$\pm$2.45 \kms, [M/H] of -0.055$\pm$0.013, and [$\alpha$/M] of 0.063$\pm$0.008 (the most $\alpha$-poor fiber position in our sample).  Any further interpretation of these results is beyond the scope of this paper.

\subsection{Abundance Gradients \label{subsec:gradients}}
\begin{figure}
\includegraphics[width=3.35in]{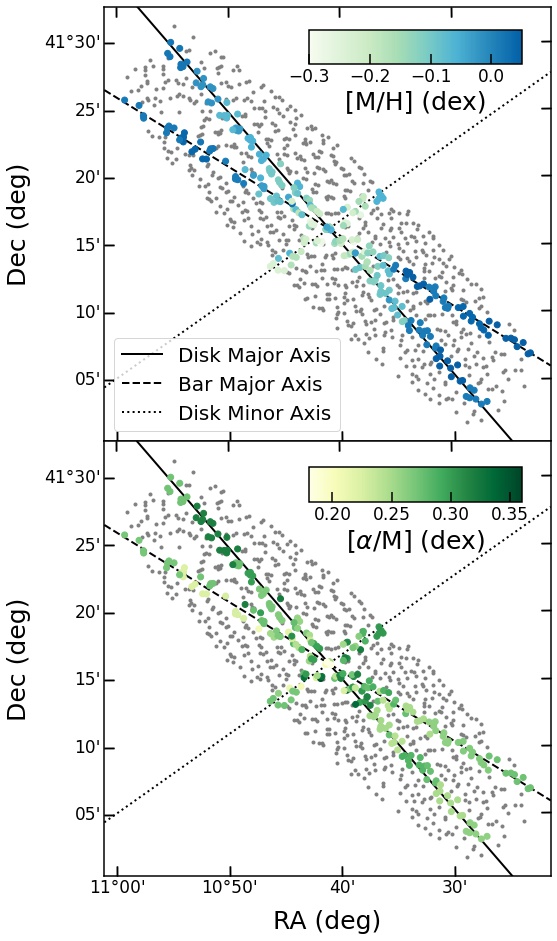}
\caption{\label{fig:gradmaps}This figure demonstrates how the gradients in Figure \ref{fig:grads} are calculated.  Fiber positions along each axis are assigned the abundance for their bin, then the radius for each fiber position is calculated.  The radius for the bin is the mean radius for fiber positions in that bin along that axis.  Unused fiber positions are denoted as gray dots.}
\end{figure}

\begin{figure*}
\includegraphics[width=\textwidth]{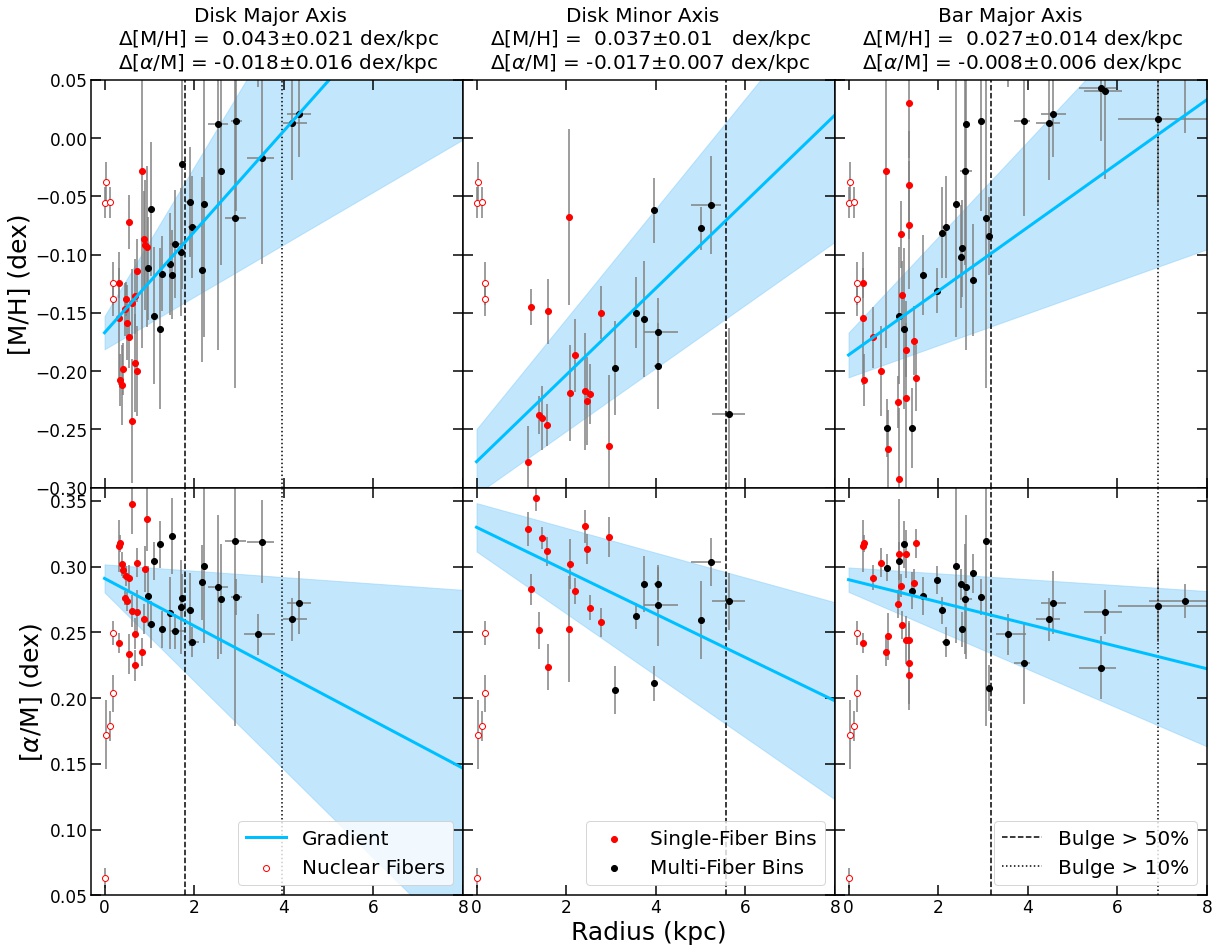}
\caption{\label{fig:grads}Metallicity (top) and $\alpha$ abundance (bottom) gradients for the fiber positions shown in Figure \ref{fig:gradmaps}.  The calculated gradients (blue lines) are the slopes of a straight line fit to the data taking abundance errors into account and are quoted in the title of each panel.  Red open circles are the five nuclear fiber positions, which are not included in gradient calculations.  Red circles are Voronoi bins with only one fiber position, black circles are Voronoi bins with more than one fiber position.  The radius error bars correspond to the minimum and maximum radius of fiber positions along that axis in each bin, and the abundance errors were determined by the jackknife procedure in Section \ref{subsec:errorbars}.  The shaded blue regions show the errors on our gradient fits.  The dashed lines show the region along each axis where the bulge makes up \textgreater50\% of the total light, dotted lines indicate this region for \textgreater90\%.}
\end{figure*}

We derived gradients for the metallicity and $\alpha$ abundance along three axes in the bulge: the disk major and minor axes, as well as the bar major axes (Figure \ref{fig:grads}).  Gradients are calculated by assigning each fiber position the abundance calculated for its bin.  We isolate the fiber positions within .01 degrees on-sky of each axis (Figure \ref{fig:gradmaps}) and calculate the average radius for these fiber positions.  This radius is used to derive the gradient by fitting a straight line to the plot of abundance vs. radius.  We only fit to fiber positions falling within the bulge region.  The central fiber positions (red open circles in Figure \ref{fig:grads}) are located in the nuclear regions of M31 (see Section \ref{subsec:nucleus}) which is made up of different stellar populations than the bulge and inner disk.  These five fiber positions are excluded from the gradient calculations.\\
\indent We find strongly positive metallicity gradients along the disk and bar major axes (0.043 and 0.027 dex/kpc respectively) with a flattening beyond $\sim$3 kpc around solar metallicity.  The spread in abundance for red points (corresponding to single-fiber-position bins) is likely stochastic in nature (See Section \ref{subsec:choices}).  The gradient along the disk minor axis is noisy but still generally positive (0.037 dex/kpc).  There is a notable decrease in  metallicity of $\sim$0.1 dex from the nuclear fiber positions to bins between 1-3 kpc, corresponding to the metallicity desert perpendicular to the bar.\\
\indent Gradients for the $\alpha$ abundance are much more scattered than those for metallicity.  They are negative, at $\sim$-0.017 dex/kpc along the disk major and minor axes and -0.0085 dex/kpc along the bar.  The single-fiber-position bins (generally associated with the bulge) show higher maximum enhancement than the multi-fiber-position bins in the disk, indicating overall enhancement of the bulge relative to the disk.  The effects of stochasticity are still apparent.  The bottom right panel of Figure \ref{fig:grads} clearly shows the asymmetry along the bar, which affects the gradient calculated here, decreasing its magnitude.\\
\indent Our abundance gradients in the bulge are likely the result of the mixing of a metal-poor, $\alpha$-rich bulge and a metal-rich, $\alpha$-poor disk.  Light in the central regions will be dominated by the bulge, but the relative fraction of this components contribution to the total light will decrease steadily with distance.  As a result, the mean metallicity gradients are steeply positive, but the intrinsic gradients for each component (classical bulge, b/p bulge, inner disk, etc.) may not be.


\subsubsection{Bulge-Disk Decomposition \label{subsubsec:decomposition}}
\begin{figure}
\includegraphics[width=3.35in]{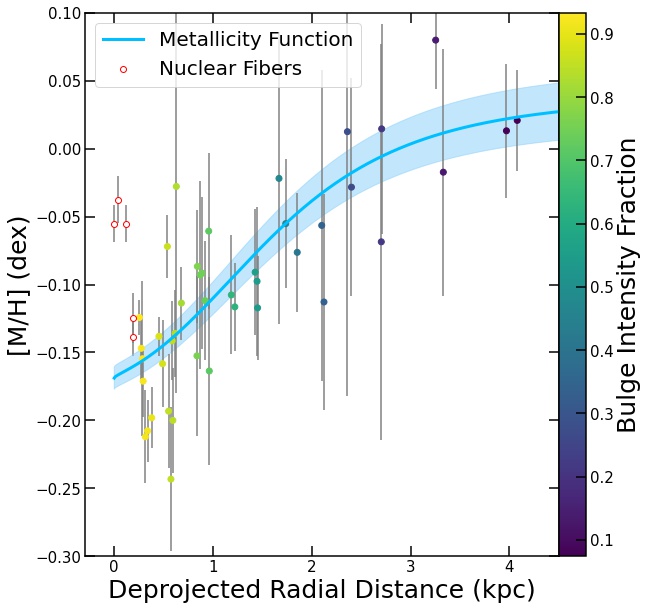}
\caption{\label{fig:decomp}Decomposition of the bulge metallicity gradient along the disk major axis.  Points were selected in the same as the top left panel of Figure \ref{fig:grads}.  Nuclear fiber positions are not included in the fit.  All other points are colored by the fraction of their light coming from the bulge.  The blue line shows how our modeled metallicity changes as a function of radius.}
\end{figure}

If we assume that the metallicity gradient in the disk is the result of the superposition of a metal-poor bulge and metal-rich disk, then we can use the photometric bulge/disk decomposition from \citep{Courteau+2011} to determine the uniform metallicity of the bulge and inner disk.  We use their Equations 2 and 4 to determine the fractional intensity of light coming from the bulge as a function of radius.  We fit their Sersic bulge and exponential disk model to our data and find our data are well represented by a bulge with [M/H] = -0.219$\pm$0.008 and a disk with [M/H] = 0.0584$\pm$0.025.  This relationship along the disk major axis is shown in Figure \ref{fig:decomp}.

\subsection{Comparison to Literature \label{subsec:litcomp}}
\subsubsection{Chemodynamics \label{subsubsec:litchemodynamics}}
The most similar analysis to ours found in the literature is from \citetalias{Opitsch+2018} (kinematics) and \citetalias{Saglia+2018} (abundances and ages).  As mentioned in Section \ref{subsec:opitsch}, these data are from the VIRUS-W optical IFU.  These IFU data are much more densely packed in the bulge than our APOGEE data, but their results are still subject to stochastic effects.  We calculated the difference between abundance in neighboring bins for all of the bins in their central region, excluding the spokes.  The standard deviation of this distribution was $\sigma$[Z/H] = 0.136 and $\sigma$[$\alpha$/M] = 0.108.  Despite this, their spatial sampling was dense enough to uncover detailed spatial structure.\\
\indent \citetalias{Opitsch+2018} performed full spectrum fitting using \textsc{ppxf} with 41 kinematic standard stars and constructed detailed kinematic maps of this region and found strong evidence for a bar from cuts of the kinematics through the disk major axis.  \citetalias{Saglia+2018} examined the same data using Lick indices to determine the chemistry of M31.  They found the inner 2.3 kpc (where the light is dominated by stars in the classical bulge) to be old (12.9 Gyr), metal-rich ([M/H] up to .35, averaging to be slightly super-solar) and $\alpha$ enhanced ([$\alpha$/Fe] of .28).  The b/p bulge is the same age at 13.1 Gyr, but is slightly sub solar with [M/H] = -.04 and less $\alpha$ abundant at [$\alpha$/Fe] = .25.  The bar is slightly younger, but has similar abundance to the classical bulge, and is metal-rich.\\
\indent In general, our results agree with this analysis.  Our kinematics show strong evidence of bar effects in both the `s' shaped velocity curve and the velocity vs. position angle plot. The spatial structure we find in our abundances is also correlated with those found in the above.  They see enhancement in metallicity along the spurs of their data that extend into the disk along the disk major axis, and depression for the spurs along the disk minor axis, much as we do.  We also find the metallicity deserts on either side of the bar, though we hesitate to identify this structure as an enhancement along the bar.  Our data are too sparsely sampled spatially, and as such are too subject to stochastic effects, to say this for sure (see Section \ref{subsec:choices} for further discussion).  \citetalias{Saglia+2018} notes near-uniform enhancement in $\alpha$ abundance in the bulge with the exception of the nuclear regions, much as we do (excepting the statistical fluctuations), and see a reduction in abundance along the north-east side of the bar.  They also find the bulge to be older than 12 Gyr.\\

\subsubsection{Gradients}
\indent Many studies have analyzed the metallicity gradient in the disk of M31.  Studies of HII regions in the disk of M31 ($\sim$4-20 kpc) have indicated a slightly negative metallicity gradient of -0.023$\pm$0.002 dex/kpc \citep{Zurita+Bresolin2012}.  Analysis of planetary nebulae in M31's outer disk show a more shallow metallicity gradient around -.011 dex/kpc from 18-43 kpc \citep{Kwitter+2012}.  The PHAT survey shows a negative metallicity gradient from 4-20 kpc that matches that of the HII analysis at -0.020$\pm$0.004 dex/kpc \citep{PHAT_metal}.  The gradients found by these studies are consistent with simulations of the merger history of M31 that indicate it had a recent major interaction, potentially with M32, 2-4 Gyr ago \citep[e.g.,][]{Hammer+2018,Dsouza+Bell2018}.\\
\indent These studies have analyzed and drawn conclusions from metallicity gradients in the disk \greaterthan 4 kpc from the center of M31.  Gradients inside this radius may result from different physical processes, as they are inside a different structure.  We do not interpret our positive gradients in the context of M31's disk formation or merger history.  Instead, as was shown in Section \ref{subsubsec:decomposition}, they can be modeled as an effect of the changing light contribution from the bulge and disk structural components.\\
\indent The gradient in the bulge was studied by \citetalias{Saglia+2018}, who calculate metallicity and $\alpha$ abundance gradients in the central 500" (roughly 1.9 kpc).  They call the inner $\sim$78" ($\sim$0.3 kpc) the classical bulge and calculate steeply negative and shallow positive gradients for [Z/H] and [$\alpha$/Fe] respectively.  \citetalias{Saglia+2018} find a shallow positive metallicity gradient out to 500" for the bar and a shallow negative metallicity gradient off the bar.  Both on- and off-bar regions show a shallow negative $\alpha$ abundance gradient.\\
\indent \citet{PHAT_metal} quotes their gradient for the outer disk, but note that their results within 4 kpc show a positive metallicity gradient closer to our results (see their Figure 4, note that their results in this region are subject to crowding and completeness effects).  Outside of 4 kpc, our results show a roughly flat gradient out to 7 kpc (see the top right panel of Figure \ref{fig:grads} for the best example), much more in line with previous studies.

\subsubsection{Bar and Bulge Properties \label{subsubsec:barbulge}}
\indent The twist in the velocity field along the disk minor axis seen in Figure \ref{fig:loessvel} is often called non-axisymmetric motion or a non-circular velocity field.  It is seen in perhaps 20-50\% of disk galaxies, most often in the gas motions \citep{Haynes+1998,Swaters+1999,Kornreich+2000,Andersen+Bershady2013,Bloom+2017}, and identified as early as 1978 \citep[e.g.,][]{Peterson+1978,Sanders+Tubbs1980}.  The strongest deviations from circular gas motions in CALIFA \citep{CALIFA} galaxies are associated with the presence of a bar \citep{Garcia-Lorenzo+2015}.  The most detailed analysis of asymmetric stellar motions was done by \citet{Stark+2018}.  They investigate $\sim$2800 MaNGA galaxies and find this feature (which they call ``inner bends") in $\sim$20-25\% of their sample.  They find it is most closely associated with having a strong bar, however roughly 45\% of the strongly barred galaxies in their sample show no distortion.  They note these numbers may be skewed by the difficulty of visually identifying bars.\\ 
\indent Our analysis finds inconclusive evidence of the metal-rich bar seen in \citetalias{Saglia+2018}.  However our results may line up well with the stellar chemodynamics of the MW's bar and bulge.  \citet{Rojas-Arriagada+2020} found that the MW's bulge can be split into three spatially overlapping populations at [Fe/H] = +0.32, -0.17, and -0.66 dex in order of decreasing population concentration.  They confirm the interpretation that metal-rich stars in the disk secularly evolve into the boxy/peanut bulge and argue that the intermediate metallicity population could have formed in-situ at high redshift.  The metal-poor stars could be associated with the early thick disk.  Similarly, \citet{Wegg+2019} find that bulge stars on the most bar-like orbits have the highest metallicities ([Fe/H] = 0.3) and inner-disk stars have lower metallicity ([Fe/H] = 0.03).  These results, as well as those of \citet{Queiroz+2021}, line up well with ours, but the nature of the MW bulge is still up for debate:  \citet{Bovy+2019} define the bar by its kinematics and find it to be more metal-poor and $\alpha$-rich on average compared to the inner disk.\\
\indent An N-body simulation of the bar's evolution from \citet{Fragkoudi+2018} indicate the bar is made up of metal-rich ([Fe/H] \textgreater 0 and metal-poor ([Fe/H] \textless 0) populations whose fractional contribution to the bar varies spatially.  The metal-rich population makes a longer, more narrow bar, whereas the metal-poor population is shorter and wider.  This results in a non-uniformly enhanced bar that is metal-poor in the center and metal-rich at the ansae.  The regions perpendicular to the bar are also metal-poor.  This final result may line up with ours best, as we find the non-nuclear fiber positions in the center to be generally the most metal poor, as well as a positive radial metallicity gradient along the bar.\\

\subsection{Examining Robustness to Analysis Choices \label{subsec:choices}}
Above, we have presented an analysis whose details depend on the methodological choices outlined in Section 3, which we will call our default method.  To test our robustness to these methodological choices, we re-analyzed our data in several different ways.  In this section we discuss why we chose our default method and the robustness of the features seen in our results to these methodological choices.

\subsubsection{Criteria for our Presented Results}
Our primary focuses when choosing which analysis variation to present were to ensure that results were consistent between neighboring bins and that gradients were reasonably tight.\\
\indent To quantify bin-to-bin consistency we calculated the difference between abundances in neighboring bins (as in Section \ref{subsubsec:litchemodynamics} for \citetalias{Saglia+2018}) and determined the standard deviation of the distribution.  For our default method, we find $\sigma$[M/H] = 0.083, and $\sigma$[$\alpha$/M] = 0.055.  This was similar to or less than our results for the other analysis variations described in the following subsections, which varied from 0.1 to 0.164 in metallicity and 0.036 to 0.06 in $\alpha$ abundance.\\
\indent To determine the tightness of gradients we calculated the residuals from our gradient fits along the disk major axis for each analysis variation.  The standard deviation of our residuals for our default method are $\sigma$[M/H] = 0.0455 and $\sigma$[$\alpha$/M] = 0.031.  These numbers for the other variations ranged from 0.046 to 0.177 in metallicity and 0.026 to 0.046 in $\alpha$ abundance.  Our presented results have the best combination of these statistics. 

\subsubsection{Binning Effects}
The binning scheme used in our final results was motivated to obtain as many radially narrow bins as possible at or above an eSNR of 60 (see Section \ref{subsec:abundancebin}).  What we lose by using this routine, however, is the ability to identify more detailed azimuthal substructure in the disk, as well as symmetry along the disk minor axis.  We created a different binning scheme that maintained symmetry across both the disk major and minor axes, but sacrificed radial narrowness and eSNR.  Analysis of this binning strategy produced similar maps to Figure \ref{fig:mosaic} and gradients to Figure \ref{fig:grads}, but introduced more stochasticity into our measurements.  Bins that were next to each other azimuthally often did not have similar abundance.  This strategy also did not show an equivalent to the $\alpha$ deficiency seen on the north-east side of the bar on the south-west side.  

\subsubsection{Continuum Treatment \label{subsubsec:continuum}}
\indent The handling of the continuum shape of our spectra has an effect on the scaling of our abundance results.  There are two main ways to handle the continuum using \textsc{ppxf}.  The first one, which we use in our default method, is to manually normalize the spectrum using a running median.  The second is to use the mdegree keyword in \textsc{ppxf}, which applies a multiplicative Legendre polynomial to the model spectrum to mimic the continuum shape of the data.  \textsc{ppxf} has full control over the shape of this polynomial; mdegree simply tells \textsc{ppxf} what order polynomial to apply.  \\
\indent We tested the effects of continuum and mdegree on our results using our models and real data.  To test the models, we took the templates with metallicity between -0.6 and +0.4 and $\alpha$ abundance of 0.1 or 0.3, and added noise and broadening to simulate our observed spectra.  This was done by gaussian broadening the templates to a dispersion consistent with our results for M31, and adding noise corresponding to the uncertainty array of a random binned spectrum.  We call these templates artificially noisy templates, or ANTs.  For one test we did no modification to the continuum of the ANTs; for the other test we added in the continuum from the binned spectrum we used to add noise (for reference, our continuum calculations are described in Section \ref{subsubsec:Masking}.)  The ANTs were analyzed using values of mdegree between zero and ten.  We found we found that the best fitting value of dispersion, metallicity, or $\alpha$ abundance across all values tested did not change with changing mdegree, and got back our input results within errors.  For the ANTs with continuum added, the dispersion increased by roughly 20 \kms\space from mdegree of zero to ten, and the metallicity increased by 0.2 to 0.5 dex, depending on the input metallicity.\\
\indent We performed a similar set of tests on our data, handling continuum three different ways: continuum normalized with no multiplicative polynomial, continuum normalized with a 2nd order multiplicative polynomial, and non-continuum normalized with an mdegree of six, eight, or ten, depending on the number of pixels in the chip.\\
\indent In general, the structures seen in our results were consistent with each method, though the consistency between neighboring bins was decreased.  However we found that the scaling of our dispersion and metallicity was correlated with mdegree. The high-order multiplicative polynomial resulted in dispersions $\sim$50 \kms\space and metallicities $\sim$0.5 dex higher than the continuum normalized versions with a 2nd order polynomial.  Metallicity results were $\sim$0.7 dex higher than results with no polynomial applied.  These increased metallicities fell outside of the A-LIST grid, and our interpolated model had to extrapolate to create these templates.\\
\indent Structurally, we found that these higher-order fits resulted in a stronger signature of bar metallicity enhancement and a more uniformly $\alpha$ enhanced bulge.  However these results were also more stochastic in the disk.  Ultimately, the low-order polynomial fits were chosen as they reduce model complexity, provide more robust results, and bring our results more inline with the A-LIST parameter space and previously published kinematics and abundances.

\subsubsection{Fixing Age to 12 Gyr \label{subsubsec:fixedages}}
As stated previously, our models are not particularly sensitive to stellar age differences of a few Gyr.  We find that most of our bins end up with ages of 12 Gyr (the edge of our model grid), with a few exceptions.  We performed our primary analysis method with the restriction that every bin must be fixed to 12 Gyr and found minimal changes to our results.  There are 10 bins with ages below 11 Gyr.  Fixing the age in these bins caused a decrease in metallicity and $\alpha$ abundance in every bin.  The average change in abundance was $\Delta$[M/H] = -0.107 and $\Delta$[$\alpha$/M] = -0.039. Additionally, the majority of bins with low ages have elevated errors in metallicity, $\alpha$ abundance, and age relative to the 12 Gyr bins.

\subsubsection{Effects of Masking \label{subsubsec:misfit}}
\indent Our 85th percentile cutoff for masking out skylines as described in Section \ref{subsubsec:Masking} had the potential to mask out the deepest sections of absorption features in our spectra.  To test the effects of our masking, we ran two tests on the ANTs from Section \ref{subsubsec:continuum}.  For the first tests we performed the fits using the mask arrays from the binned spectra used to make the ANTs.  For the second test we masked out no pixels other than the misfit regions and chip gaps.  In both cases we found minimal bias (-0.01 and -0.02 dex, respectively) of our output metallicities from the input ones, indicating our results are not affected by our masking routine.\\
\indent We also found several regions in our spectra that were consistently poorly fit across the majority of our bins.  These regions fell into two categories: known skylines and poorly-modelled lines.\\
\indent We were not able to determine every skyline using the strategy outlined in Section \ref{subsubsec:Masking}, and sometimes for skylines that we did detect, we did not mask out all of the affected pixels.  To account for this, we mask out pixels within 2.5 \AA\space of known skylines prior to fitting.  We utilize the skylines from \citet{Rousselot+2000} with intensities \greaterthan 0.5.\\
\indent The three larger regions that are masked are due to poor data-model matching.  We were consistently unable to fit these regions across a majority of our spectra.  The region from 15360 to 15395 \AA\space is a deep, broad line whose depth was not well fit.  The region from 16030 to 16070 \AA\space is another deep line, parts of which were sometimes masked by our skyline identification and other times not.  Lastly, the region from 16210 to 16240 \AA\space is a series of smaller lines with inconsistent fits from spectrum to spectrum.\\

\section{Summary and Conclusions}
\indent In this paper we analyzed 963 high resolution, near infrared, integrated light APOGEE spectra of the bulge and inner disk of M31.  We spatially binned these spectra into 164 bins with an eSNR $\gtrsim$60.  These binned spectra were fit with interpolated SSP model spectral templates from A-LIST \citep{Ashok+2021}.  The result is maps of the mean radial velocity, velocity dispersion, metallicity, $\alpha$ abundance, and age of the stellar populations in the inner $\sim$7 kpc of M31 (Figure \ref{fig:mosaic}).  We also computed metallicity and $\alpha$ abundance gradients along the disk major and minor axes and the bar major axis.  A summary of our primary conclusions is below.\\
\begin{itemize}
\item[-]\textsc{Radial Velocity:} The line where the radial velocity is equal to the systemic velocity of M31 (-300 \kms) is twisted in the direction of the bar major axis in the bulge.  This feature is strongly associated with the presence of a bar \citep[e.g.,][]{Stark+2018}.
\item[-]\textsc{Velocity Dispersion:} The velocity dispersion of M31 does not peak in the very center.  Our results find a peak $\sim$1 kpc from the center to the south-east.  This is consistent with the findings of \citetalias{Opitsch+2018}, who identify a ring of increased dispersion this far from M31's center.  The dispersion decreases with distance from M31's center at a rate of $\sim$10 \kms\space per kpc.
\item[-]\textsc{Metallicity:} The metallicity of the bulge of M31 is sub-solar ([M/H] = $-0.149^{+0.067}_{-0.081}$), and the disk is nearly solar ([M/H] = $-0.023^{+0.050}_{-0.052}$).  We find regions of decreased metallicity relative to the disk on either side of the bar.  The metallicity here is sub solar, [M/H] $\sim$ -0.2.  The bar of M31 has been found to be enhanced relative to the rest of the bulge \citepalias{Saglia+2018}, but we are unable to clearly identify this pattern.
\item[-]\textsc{$\alpha$ Abundance:} The bulge of M31 is enhanced in $\alpha$ elements ([$\alpha$/M] = $0.281^{+0.035}_{-0.038}$).  Statistical fluctuations are noted in results for individual fiber positions in the bulge, but these fluctuations do not appear to trace a coherent spatial pattern.  The disk is similarly enhanced on average ([$\alpha$/M] = $0.274^{+0.020}_{-0.025}$) but has more defined structure. We find a region of decreased $\alpha$ abundance to the north-east of the bulge along the bar major axis.  This could be the end of the bar sticking out of the bulge above the disk, but further analysis is needed to interpret this feature.
\item[-]\textsc{Stellar Age:} Both the bulge and disk are uniformly old at $\ge12$ Gyr, with the exception of a small number of disparate bins.  Fixing the age of these disparate bins to 12 Gyr decreased both the metallicity and $\alpha$ abundance results for these bins to be more in line with adjacent bins.
\item[-]\textsc{Nuclear Region:} M31's nucleus contains a stellar population distinct from the bulge and the disk.  The nucleus is metal enhanced ([M/H]$\simeq$-0.05) and $\alpha$ deficient ([$\alpha$/M]$\simeq$0.18) relative to the rest of the bulge.  We find the very center of M31 is the most $\alpha$-poor of any region (.006 dex), and is moving away from the sun relative to the rest of M31, with a velocity of $-225.56^{+0.486}_{-0.459}$ \kms.
\item[-]\textsc{Abundance Gradients:} The bulge features steeply positive metallicity gradients, steepest along the disk major axis at $\Delta$[M/H]=0.043$\pm$0.021 dex/kpc.  The $\alpha$ abundance gradients in this region are negative, though shallower, and noisy.  The gradient along the disk major axis is $\Delta$[$\alpha$/M]=-0.018$\pm$0.016 dex/kpc.
\item[-]\textsc{Bulge/Disk Decomposition:} The gradients in the bulge are not necessarily intrinsic to the bulge stellar populations.  They can also be modeled as a change in light contribution from a uniformly metal-poor bulge ([M/H]=-0.219$\pm$0.008) and uniformly metal rich inner disk ([M/H]=0.0584$\pm$0.025).
\end{itemize}

\indent These results are robust to a number of different methodological choices.  Our findings agree well with those of previous resolved and integrated light studies of M31 and are consistent with our understanding of the MW bulge and inner disk.  The data used to produce the maps in Figure \ref{fig:mosaic} are published with this work as formatted in Table \ref{tab:result} in Appendix \ref{sec:tables}.\\
\indent Future work will reanalyze these spectra using an expanded methodology that will fit two spectral templates to the data in order to characterize multiple kinematically distinct co-spatial stellar populations in M31.  Such populations could include the separate classical and boxy/peanut shaped components in the bulge and the thick and thin disks, with a particular emphasis on identifying a trend like the so-called $\alpha$-bimodality that is seen in the MW \citep[e.g.,][]{Haywood+2013, Hayden+2015}.\\

\begin{center}
ACKNOWLEDGEMENTS
\end{center}

A very heartfelt thanks is extended to the APO plate pluggers who figured out how plug our atypical plate designs (see Figure \ref{fig:plugplate}) - their work was vital to this project.  The authors thank Nicholas Boardman and Jianhui Lian for their helpful discussions on this work.  BJG, GZ, and AS acknowledge support for this project was provided by NSF grant AST-1911129.

Funding for the Sloan Digital Sky Survey IV has been provided by the Alfred P. Sloan Foundation, the U.S. Department of Energy Office of Science, and the Participating Institutions. SDSS-IV acknowledges
support and resources from the Center for High-Performance Computing at
the University of Utah. The SDSS web site is www.sdss.org.

SDSS-IV is managed by the Astrophysical Research Consortium for the 
Participating Institutions of the SDSS Collaboration including the 
Brazilian Participation Group, the Carnegie Institution for Science, 
Carnegie Mellon University, the Chilean Participation Group, the French Participation Group, Harvard-Smithsonian Center for Astrophysics, 
Instituto de Astrof\'isica de Canarias, The Johns Hopkins University, Kavli Institute for the Physics and Mathematics of the Universe (IPMU) / 
University of Tokyo, the Korean Participation Group, Lawrence Berkeley National Laboratory, 
Leibniz Institut f\"ur Astrophysik Potsdam (AIP),  
Max-Planck-Institut f\"ur Astronomie (MPIA Heidelberg), 
Max-Planck-Institut f\"ur Astrophysik (MPA Garching), 
Max-Planck-Institut f\"ur Extraterrestrische Physik (MPE), 
National Astronomical Observatories of China, New Mexico State University, 
New York University, University of Notre Dame, 
Observat\'ario Nacional / MCTI, The Ohio State University, 
Pennsylvania State University, Shanghai Astronomical Observatory, 
United Kingdom Participation Group,
Universidad Nacional Aut\'onoma de M\'exico, University of Arizona, 
University of Colorado Boulder, University of Oxford, University of Portsmouth, 
University of Utah, University of Virginia, University of Washington, University of Wisconsin, 
Vanderbilt University, and Yale University.

The Digitized Sky Survey was produced at the Space Telescope Science Institute under U.S. Government grant NAG W-2166. The images of these surveys are based on photographic data obtained using the Oschin Schmidt Telescope on Palomar Mountain and the UK Schmidt Telescope. The plates were processed into the present compressed digital form with the permission of these institutions.

\bibliographystyle{aasjournal}
\bibliography{main}

\appendix
\section{Data Tables \label{sec:tables}}
The results of our analysis will be published in electronic data tables taking the same form as Table \ref{tab:result}.  These tables can be used to create maps similar to those in Figure \ref{fig:mosaic}.
\begin{splitdeluxetable*}{ccccccccccccBllllllllll}
\tablenum{1}
\tablecaption{Description of bins and chemodynamics results from our analysis\label{tab:result}}
\tablehead{\colhead{Bin} & \colhead{\# of Fibers} & \multicolumn{2}{c}{------RA------} & \multicolumn{2}{c}{------Dec------} & \multicolumn{2}{c}{------Radius------} & \multicolumn{2}{c}{---Position Angle---} & \colhead{$\chi^2$} & \colhead{eSNR} & \multicolumn{2}{c}{------Velocity------} & \multicolumn{2}{c}{---Dispersion---} & \multicolumn{2}{c}{------[M/H]------} & \multicolumn{2}{c}{------[$\alpha$/M]------} & \multicolumn{2}{c}{------Age------}\\
\colhead{BIN} & \colhead{NFib} & \colhead{MinRA} & \colhead{MaxRA} & \colhead{MinDec} & \colhead{MaxDec} & \colhead{MinRad} & \colhead{MaxRad} & \colhead{MinPos} & \colhead{MaxPos} & \colhead{chi2} & \colhead{eSNR} & \colhead{v} & \colhead{v\_err} & \colhead{d} & \colhead{d\_err} & \colhead{m} & \colhead{m\_err} & \colhead{a} & \colhead{a\_err} & \colhead{A} & \colhead{A\_err}\\
\colhead{} & \colhead{} & \colhead{deg} & \colhead{deg} & \colhead{deg} & \colhead{deg} & \colhead{kpc} & \colhead{kpc} & \colhead{deg} & \colhead{deg} & \colhead{} & \colhead{} & \colhead{\kms} & \colhead{\kms} & \colhead{\kms} & \colhead{\kms} & \colhead{dex} & \colhead{dex} & \colhead{dex} & \colhead{dex} & \colhead{Gyr} & \colhead{Gyr}}
\startdata
1   & 1   & 10.685 & 10.685 & 41.269 & 41.269 & 0.0   & 0.0   & 0.0     & 0.0     & 12.251 & 419.807 & -225.559 & 3.235  & 151.913 & 2.449 & -0.055 & 0.013 & 0.063 & 0.008 & 11.988 & 0.006 \\
11  & 1   & 10.645 & 10.645 & 41.239 & 41.239 & 0.832 & 0.832 & 186.791 & 186.791 & 1.908  & 71.253  & -368.24  & 10.809 & 122.906 & 5.11  & -0.028 & 0.152 & 0.235 & 0.011 & 11.926 & 0.024 \\
15  & 1   & 10.65  & 10.65  & 41.218 & 41.218 & 1.056 & 1.056 & 169.237 & 169.237 & 1.553  & 50.449  & -399.717 & 8.577  & 133.01  & 4.683 & -0.158 & 0.069 & 0.352 & 0.025 & 7.425  & 4.283 \\
37  & 1   & 10.639 & 10.639 & 41.264 & 41.264 & 1.954 & 1.954 & 224.39  & 224.39  & 2.466  & 103.435 & -337.403 & 2.498  & 139.653 & 4.221 & -0.23  & 0.049 & 0.305 & 0.011 & 11.924 & 0.024 \\
57  & 101 & 10.427 & 10.675 & 41.03  & 41.27  & 3.578 & 5.218 & 145.944 & 232.709 & 3.471  & 74.998  & -443.396 & nan    & 83.272  & 5.09  & 0.013  & 0.049 & 0.26  & 0.016 & 11.971 & 0.059 \\
69  & 14  & 10.603 & 10.65  & 41.303 & 41.356 & 4.781 & 5.56  & 263.392 & 305.275 & 2.449  & 79.582  & -280.668 & nan    & 92.891  & 4.041 & -0.058 & 0.042 & 0.304 & 0.018 & 11.962 & 0.024 \\
72  & 48  & 10.386 & 10.568 & 41.073 & 41.29  & 6.101 & 8.106 & 189.877 & 245.273 & 2.566  & 58.017  & -410.188 & nan    & 82.715  & 5.449 & 0.136  & 0.132 & 0.274 & 0.013 & 12.0   & nan   \\
78  & 2   & 10.656 & 10.66  & 41.317 & 41.317 & 3.232 & 3.385 & 297.8   & 301.01  & 1.874  & 72.098  & -271.675 & nan    & 111.532 & 1.765 & -0.086 & 0.023 & 0.268 & 0.009 & 11.954 & 0.009 \\
96  & 1   & 10.675 & 10.675 & 41.289 & 41.289 & 1.329 & 1.329 & 302.063 & 302.063 & 2.218  & 118.54  & -268.25  & 2.928  & 131.968 & 2.13  & -0.238 & 0.013 & 0.291 & 0.005 & 11.976 & 0.011 \\
123 & 1   & 10.736 & 10.736 & 41.269 & 41.269 & 2.396 & 2.396 & 52.166  & 52.166  & 1.843  & 81.905  & -274.832 & 5.836  & 131.064 & 1.118 & -0.138 & 0.092 & 0.224 & 0.031 & 11.851 & 4.779 \\
139 & 1   & 10.73  & 10.73  & 41.313 & 41.313 & 0.88  & 0.88  & 359.736 & 359.736 & 1.58   & 51.68   & -242.241 & 6.168  & 124.011 & 5.113 & -0.087 & 0.042 & 0.26  & 0.011 & 11.884 & 0.15  \\
150 & 1   & 10.706 & 10.706 & 41.294 & 41.294 & 0.473 & 0.473 & 354.708 & 354.708 & 1.985  & 99.435  & -248.027 & 5.946  & 132.642 & 4.73  & -0.138 & 0.014 & 0.293 & 0.008 & 11.967 & 0.018 \\
156 & 1   & 10.693 & 10.693 & 41.25  & 41.25  & 1.217 & 1.217 & 123.823 & 123.823 & 1.92   & 104.577 & -318.641 & 3.931  & 153.07  & 3.635 & -0.35  & 0.045 & 0.343 & 0.014 & 11.911 & 0.05  \\
164 & 6   & 10.76  & 10.776 & 41.233 & 41.253 & 4.854 & 5.27  & 65.274  & 84.65   & 2.227  & 76.315  & -296.462 & nan    & 100.632 & 4.095 & -0.077 & 0.018 & 0.259 & 0.03  & 11.975 & 0.025 \\
\\
\enddata
\tablecomments{Table \ref{tab:result} is published in its entirety in the machine-readable format. A portion is shown here for guidance regarding its form and content.\\The top row of the header describes the type of quantity given in the columns.  The middle row gives the column name in the published .csv file.  ``\_err" names correspond to the error bounds on our measurements.  Units for each quantity are given in the bottom row of the header. Bins 57, 69, 72, 78, and 164 have nans for v\_err because they are multi-fiber-position bins, and as such their velocities were determined from the average \citetalias{Opitsch+2018} velocity of all the fiber positions in each bin (see Section \ref{subsec:opitsch}).  Bin 72 has nan for A\_err because its age was fixed to 12 Gyr (see Sections \ref{subsubsec:ages} and \ref{subsubsec:fixedages}).}
\end{splitdeluxetable*}

\section{Plug Plate \label{sec:plate}}
\begin{figure}[ht!]
\includegraphics[width=\textwidth]{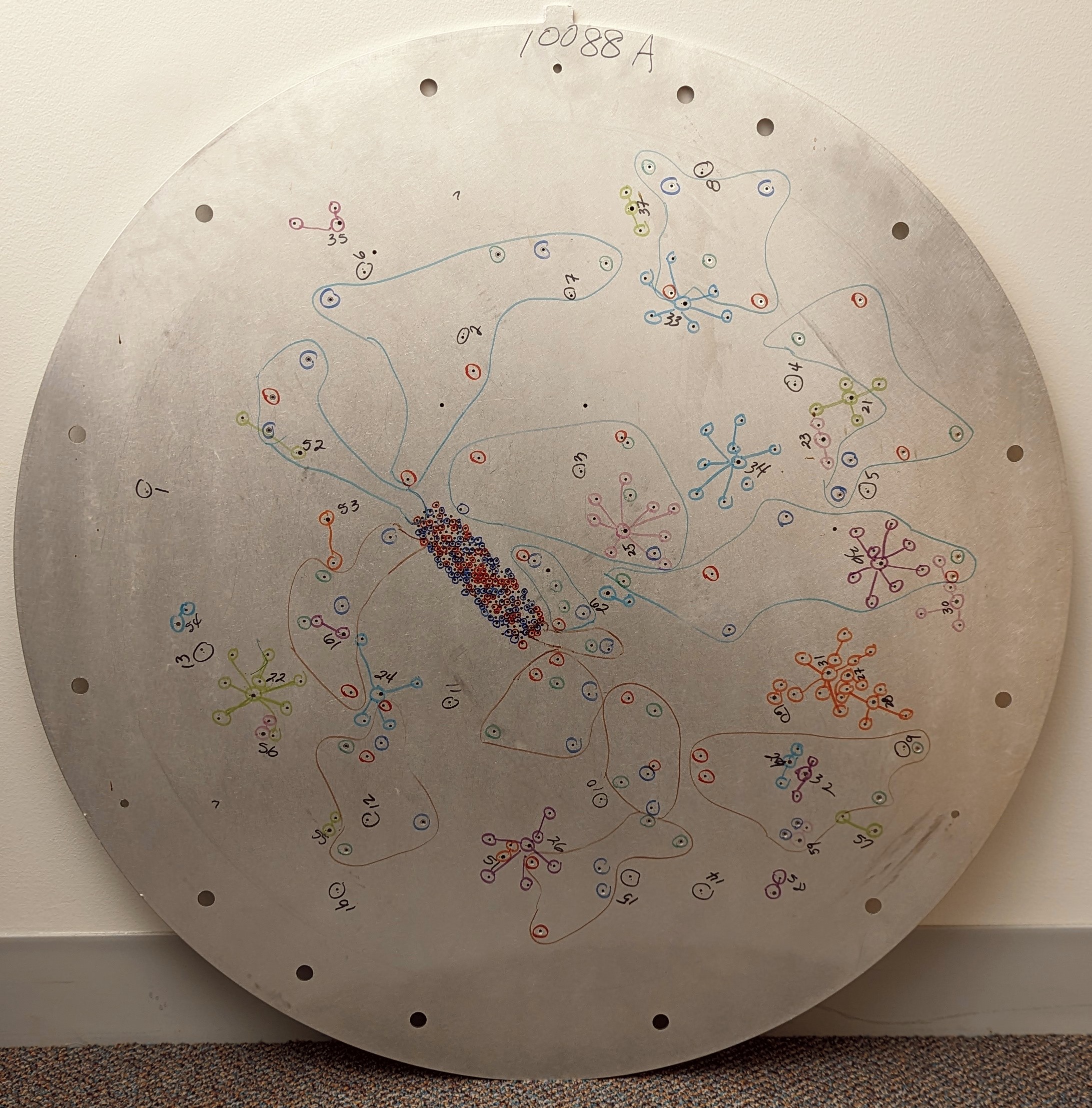}
\caption{A plug plate of one of the designs used to observe M31 with APOGEE. 
 The M31 fiber positions are all located in the densly packed rectangle just left of the center of the plate.\label{fig:plugplate}}
\end{figure}

One of the five plates designed to take observations of M31 can be seen in Figure \ref{fig:plugplate}.  This plate is unusual compared to the thousands of other APOGEE plug plates that are typically designed to observe individual stars across the whole plate, rather than a galaxy in the middle.  The M31 fiber positions are the densely packed positions just left of the center of this plate.  Other positions on this plate were used for sky and telluric flux measurements, taking spectra of M31 halo objects such as luminous blue variables or GCs \citep{Sakari+2016}, and measurements of galaxies for MaNGA \citep{MaNGA}.

\section{Comparison of 018 Velocities to Measured Velocities \label{sec:velcomp}}
We compare the \citetalias{Opitsch+2018} velocities from Section \ref{subsec:opitsch} to ones we determined independently from our data to ensure the validity of our usage of the \citetalias{Opitsch+2018} velocities in this work.  We Voronoi binned our fiber positions in the same way as Section \ref{subsec:abundancebin}, but for a target eSNR of 20, as we cannot determine reliable kinematics below this limit.  To determine kinematics, we perform the general fitting routine from Section \ref{subsec:fitting}, but run the MCMC in just two dimensions, one for velocity and the other for dispersion.  For each step in the MCMC we fit the data with the three templates with the highest weighting in the preliminary fit, much the same as in Section \ref{subsec:cutfibs}.  The result is a tight correlation between our results and the \citetalias{Opitsch+2018} velocities with a bias of -0.653 \kms\space and a scatter of $\sim$19 \kms.  Figure \ref{fig:viruscomp} shows our results for this exercise on the left and the velocities from \citetalias{Opitsch+2018} on the right.  Note the data on the right are NOT the LOESS-smoothed \citetalias{Opitsch+2018} velocities from Section \ref{subsec:opitsch}, but rather the raw data from the paper.  This figure also shows the differences in our data coverage of M31's bulge and disk.
\begin{figure}
\centering
\includegraphics[width=.82\textwidth]{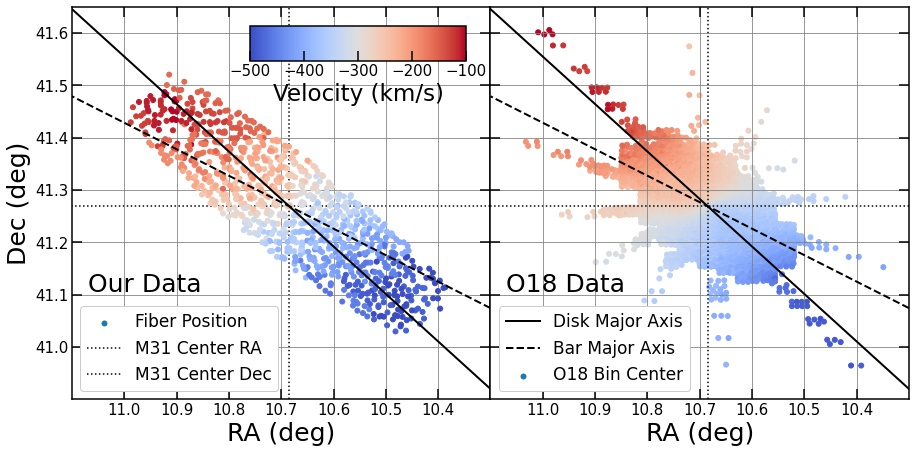}
\end{figure}
\begin{figure}
\centering
\includegraphics[width=.82\textwidth]{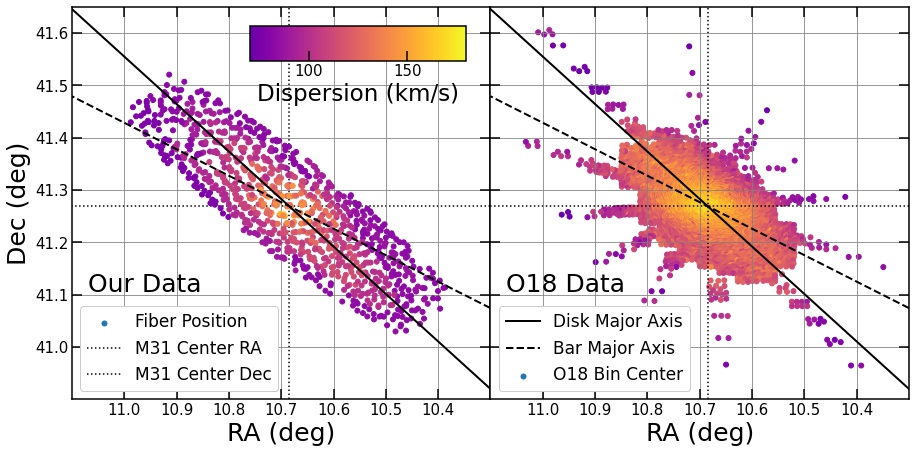}
\end{figure}
\begin{figure}
\centering
\includegraphics[width=.82\textwidth]{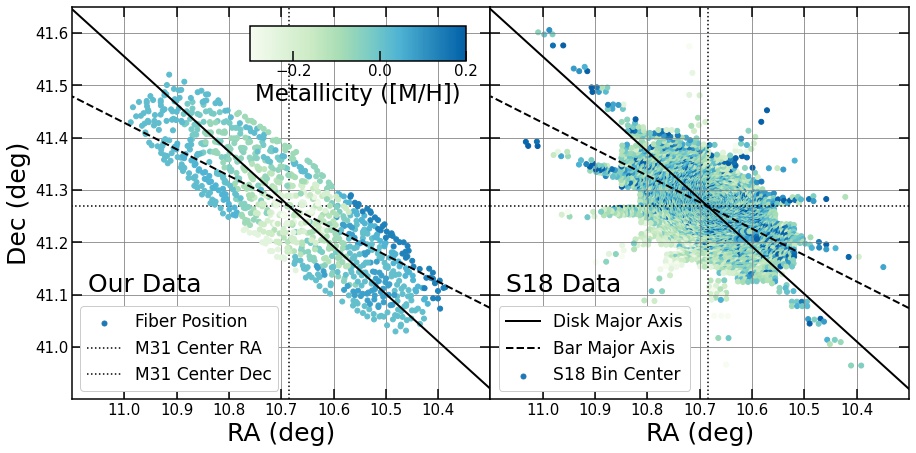}
\end{figure}
\begin{figure}
\centering
\includegraphics[width=.82\textwidth]{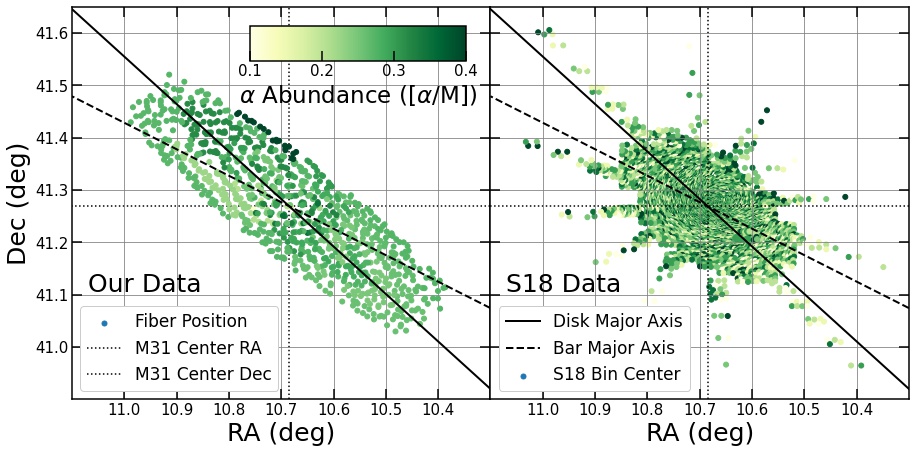}
\end{figure}
\begin{figure}
\centering
\includegraphics[width=.82\textwidth]{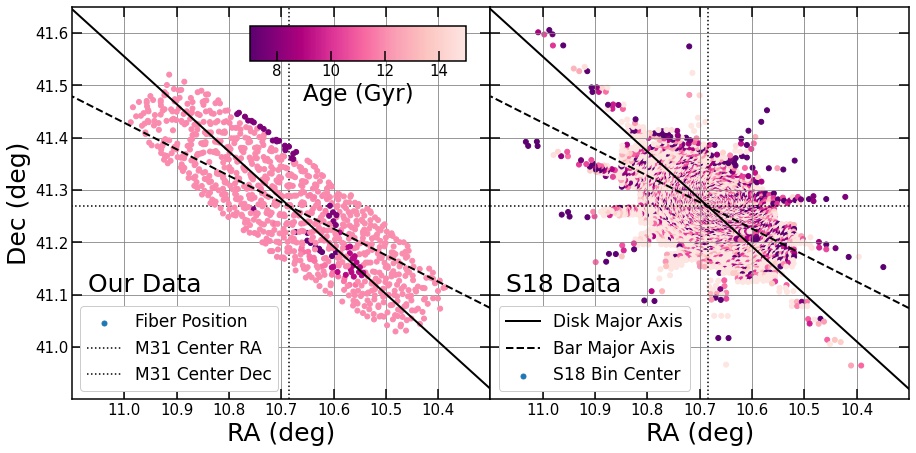}
\caption{Comparison of parameters we determined (left) to those published in \citetalias{Opitsch+2018} (right).  Note the differences in data coverage between the two data sets.  Our data spans more of the inner disk continuously opposed to the spokes in \citetalias{Opitsch+2018}'s data set. 
 The \citetalias{Opitsch+2018} data much more densely samples the bulge than ours.\label{fig:viruscomp}}
\end{figure}

\end{document}